\documentclass[a4paper]{article}
\usepackage[left=2cm,right=2cm,top=2.5cm,bottom=2cm]{geometry}
\usepackage{graphicx}
\usepackage{float}
\usepackage{amsmath,amssymb}
\usepackage{indentfirst} % 首行自动缩进
\usepackage[numbers]{natbib}
\usepackage{hyperref}
\usepackage{authblk}

\usepackage{tocloft}

\usepackage{tikz}
\usetikzlibrary{shapes.geometric, arrows.meta, positioning, calc}

\tikzset{
    mybox/.style = {
        rectangle, draw, rounded corners,
        text width=6cm, text centered,
        minimum height=1cm,
        font=\small
    }
}

\title{Gravito-Electromagnetic perturbation of higher derivative corrected Kerr-Newman black hole}

\author[1]{Yuanyuan Ji\thanks{yuanyuanji2025@163.com}}
\author[2]{Zheng-Wen Long\thanks{Corresponding author; zwlong@gzu.edu.cn}}
\affil[1]{\small College of Physics,Guizhou University,Guiyang 550025, China}

\date{}
\begin{document}
	\maketitle	
	\abstract
    We study linear gravito-electromagnetic perturbations of a slowly rotating 4-derivative-corrected Kerr–Newman (KN) black hole (BH), using a perturbative approach beyond the KN geometry up to first order in angular momentum, first order in the higher-derivative parameters $c_i,i=1,2...8$, and 14th order in the BH electric charge. We derive the corrected coupled radial master equations for both axial and polar perturbation sectors. We extend the asymptotic iteration method (AIM) to apply to coupled second-order ordinary differential equations (ODEs) and employ it to compute the quasinormal frequencies (QNFs) of the system.Our results show $10\%$ pronounced isospectrality breakings percentage between the two parity sectors, induced by 
$c_6$.This indicate the equations for the two parity sectors completely lose the transformation symmetry that relates them in the KN case.The very tiny frequency shifts induced by \(c_4\) support the interpretation that this parameter represents a redundant degree of freedom in the effective field theory.Finally, we parameterize the QNFs via a analytical fitting procedure for broader applicability.
\newpage
% --- 生成目录 ---
\tableofcontents

	\section{Introduction}
    \label{intro}

Kerr-Newman black hole represents the most general axisymmetric solution of Einstein-Maxwell theory in general relativity (GR), characterized by three parameters---mass $M$, charge $Q$, and angular momentum $a$---which correspond precisely to the three ``hairs'' allowed by the no-hair theorem \cite{Israel1967,Carter1971,Robinson1975}. Due to the coupling between the gravitational and electromagnetic fields, the perturbation equations in this background cannot be decoupled into separate single master equation for individual fields. Consequently, the computation of quasinormal modes (QNMs) for rapidly rotating Kerr-Newman BHs has long remained a challenging problem, until Dias\cite{Dias:2015wqa} use fully numerical approaches succeeded in obtaining QNM frequencies for arbitrary spin parameters.

As the intermediate signal in gravitational-wave observations, QNMs serve as a powerful probe of the source properties, since their frequencies and damping times are determined entirely by geometric parameters of the underlying compact object, independent of the initial perturbation \cite{Kokkotas1999,Berti2009}. Therefore, accurate QNM studies not only allow tests of general relativity but also offer a promising avenue for discovering new physics beyond the standard paradigm.

Due to the highly nonlinear and self-interaction of gravitational field equations,complete analytical metric solutions are available mostly for a few highly symmetric models.The rotating gravitational system beyond GR with less symmetry and additional degrees of freedom would present much mathematical challenge.To investigate gravitational perturbations of such general spinning BHs, one may resort to fully numerical schemes like spectral methods\cite{Chung:2024vaf}, which are particularly suitable for exploring the full parameter space of modified gravity, rapidly rotating BH---especially the near-extremal ones. Alternatively, analytical treatments often employ perturbative expansions in a small parameter to approximate the dynamics. These include curvature-perturbation approaches within the Newman--Penrose formalism\cite{Li:2022pcy} and direct metric-perturbation \cite{Srivastava:2021imr}. In particular, for the first kind,Ref\cite{Li:2022pcy,Cano:2023tmv} established a universal Teukolsky equation,which was subsequently apply to the study of gravitational perturbations of high rotating vaccum spacetime in the high-derivative gravity theory\cite{Cano:2023jbk}.Unfortunately, this elegant formalism is not directly applicable to our model, which is non-vacuum and retains coupled field equations. We therefore resort to the metric-perturbation scheme at linear order in the angular momentum to explore this system.

Chandrasekhar has proved the isospectrality (ISO) of perturbations of the Schwarzschild BH: the radial perturbation equations for the two parity can be mapped onto each other via the Chandrasekhar transformation\cite{Chandrasekhar1975,ChandrasekharDetweiler1975}.But the ISO may be broken in nonzero cosmological constant spacetimes or in theories beyond GR \cite{Cardoso:2018ptl,Blazquez-Salcedo:2016enn,Blazquez-Salcedo:2017txk}.Isospectrality breaking has been extensively studied in various settings. Ref.~\cite{Brito:2018hjh} investigated QNMs ISO of weakly charged Einstein--Maxwell--dilaton BHs (including rotating ones via a modified Dudley--Finley approximation). Refs.~\cite{Toshmatov:2018tyo,Toshmatov:2018ell} studied pure EM perturbations of frozen-geometry BHs in GR coupled to nonlinear electrodynamics.Ref.~\cite{Nomura:2021efi} examined spherically symmetric charged BHs in nonlinear electrodynamics, with only a brief estimate of curvature--EM coupling effects and a focus on pure nonlinear electrodynamics contributions on isospectrality. Ref.~\cite{Boyce:2025fpr} analyzed the corrected RN black hole in EFT (including the extremal case) but omitted rotation.
Our work focuses on gravito-electromagnetic perturbations of a highly charged, slowly rotating black hole in EFT,The study of which, to our knowledge remain unexplored,we takes an initial step toward filling this gap,aiming to explore the new physics that higher-derivative corrections may bring. and we obtain new results on isospectrality breaking in this regime.

The remainder of this paper is organized as follows. In Sect.~2, we introduce the four-derivative extension beyond the Einstein--Maxwell theory. In Sect.~3, we briefly describe the approximate metric solution solving procedure and re-evaluate the solution to 14th order in electric charge $Q$. In Sect.~4, we present the perturbation scheme,and leave the the radial-angular decouple process review in appendix A.1. In Sect.~6,we describe the reduction procedure and derived the corrected coupled radial master equations for both axial and polar sectors---this constitutes a central result of our work. In Sect.~7, we extend the asymptotic iteration method (AIM) to apply to two coupled second-order ODEs, and use it to calculate the QNFs of this system. In Sect.~8, we display the QNFs results we derived. In Sect.~9, we discuss about the whole work in which we use the natural units $G=c=1$.

    \section{EFT extension
of Einstein-Maxwell theory}
Dispite a series of successes in physical experiment and observation tests, there remain unsolved issues about GR since its inception ,one of which is the quantization difficulty with non-renormalizability and ultraviolet divergence.In response,numerous modified theories have been proposed,including the Effective field theory (EFT)\cite{Endlich:2017tqa,Cardoso:2018ptl, Codello:2015wua},which suggests that general relativity should emerge as the low-energy effective field limit of quantum gravity. Based on a few widely accepted physical principles,it provides a general form of corrections to GR with the action expanded as a power series of higher order curvature,that are suppressed by ultraviolet cutoff scales.Ref.~\cite{Hu:2023qhs} extended this higher-derivative framework to the Einstein--Maxwell system,which contains both even- and odd-parity terms. Since we are concerned only with the leading-order corrections, the two parity sectors can be studied separately.In this work, we focus on the even-parity sector of this four-derivative extension theory in four-dimensional spacetime, while leaving the odd-parity sector for future investigation. The even-parity part of the action is given by
\begin{align}\label{eq:1}
S &= \int d^4 x\sqrt{-g}\frac{1}{2\kappa^2} ( \mathcal{L}_{EM}+  \mathcal{L}_{HD})  \\
\mathcal{L}_{EM} &=R - \frac{1}{4} F_{\mu \nu}F^{\mu \nu}\\
\mathcal{L}_{HD} &= c_1R^2 +c_2R_{\mu \nu}R^{\mu \nu} + c_3R_{\mu \nu \gamma \kappa}R^{\mu \nu \gamma \kappa} +c_4RF^2 + c_5R_{\mu \nu}F^{\mu \gamma}F_{\gamma}^{\nu} + c_6R_{\mu \nu \gamma \kappa}F^{\mu \nu}F^{\gamma \kappa} \nonumber\\
&\quad +c_7(F^2)^2 + c_8F_{\mu \nu}F^{\nu \gamma}F_{\gamma \kappa}F^{\kappa \mu}  
\nonumber\end{align}
where \(F_{\mu\nu} \equiv \partial_\mu A_\nu - \partial_\nu A_\mu\) is the Maxwell field strength, with \(A_\mu\) denoting the electromagnetic potential. The parameters \(c_i\) are the higher-derivative couplings encode all the information of this extension theory. Here, \(g\), \(R\), \(R_{\mu\nu}\), and \(R_{\mu\nu\gamma\kappa}\) denote the determinant of the metric, the Ricci scalar, the Ricci tensor, and the Riemann curvature tensor, respectively. The first term in the action correspond to the Einstein--Maxwell system in GR, while \(\mathcal{L}_{\rm HD}\) is the higher-derivative correction, which is composed of higher-order curvature terms, nonlinear electromagnetic terms, and curvature--electromagnetic coupling terms. A natural question is which of these contributions induces the largest isospectral splitting and what other signatures beyond GR they may produce. Varying the action with respect to \(g_{\mu\nu}\) and \(A_\mu\) yields the gravitational and Maxwell field equations of the theory, which are given by
\begin{align}\label{eq:3}
&R_{\mu \nu} - \frac{1}{2} Rg_{\mu \nu} = \frac{1}{2} (T^{\rm \small (EM)}_{\mu \nu} + T^{\rm \small (HD)}_{\mu \nu})
\end{align}
\begin{align}\label{eq:4}
&\nabla_{\mu}F^{\mu \nu} =  J_{\rm \small HD}^{\nu},
\end{align}
where \(T_{\mu \nu} = F_{\mu}^{\gamma}F_{\nu \gamma} - \frac{1}{4} g_{\mu \nu}F_{\gamma \kappa}F^{\gamma \kappa}\) is the energy-momentum tensor of Maxwell field. The effective energy-momentum tensor $\Delta T_{\mu \nu}$ and electric current \(\Delta J^{\nu}\) is given by
\begin{align}
T^{\rm \small (HD)}_{\mu \nu} &= \frac{-2}{\sqrt{-g}}\frac{\delta(\sqrt{-g}\mathcal{L}_{HD})}{\delta g^{\mu \nu}} = -2\Bigl( P^{\alpha \kappa \gamma}_{(\mu} R_{\nu)\alpha \kappa \gamma} -2\nabla^{\gamma}\nabla^{\kappa}P_{\gamma(\mu \nu)\kappa} -\frac{1}{2}\mathcal{L}_{HD}g_{\mu \nu} + \frac{1}{2}M^{\alpha}_{(\mu} F_{\nu)\alpha} \Bigr) \label{eq:5} \\
P^{\mu \nu \gamma \kappa} &= \frac{\partial\mathcal{L}_{HD}}{\partial R_{\mu \nu \gamma \kappa}} \nonumber\\
&= 2c_1R g^{\mu[\gamma}g^{\kappa]\nu} + c_2(R^{\mu[\gamma}g^{\kappa]\nu} - R^{\nu[\gamma}g^{\kappa]\mu}) + 2c_3R^{\mu \nu \gamma \kappa} \nonumber\\
&\quad + c_4g^{\mu[\gamma}g^{\kappa]\nu}F^2 + c_5\frac{1}{2}(S^{\mu[\gamma}g^{\kappa]\nu} - S^{\nu[\gamma}g^{\kappa]\mu}) + c_6F^{\mu \nu}F^{\gamma \kappa}  \label{eq:6}
\end{align}

where \(S^{\mu \nu} = F^{\mu}_{\gamma}F^{\nu \gamma}\).and
\begin{align}
J_{\rm \small HD}^{\nu} = \nabla_{\mu}M^{\mu \nu}
\end{align}
\begin{align}
M^{\mu \nu} &= 2\frac{\partial\mathcal{L}_{HD}}{\partial F_{\mu \nu}} = 4(c_{4}R F^{\mu \nu} - c_{5}R_{[\gamma}^{\mu}F^{\nu ]\gamma} + c_{6}R^{\mu \nu \gamma \kappa}F_{\gamma \kappa} + 2c_{7}F^{\mu \nu}F^2 + 2c_{8}F^{\mu \gamma}F^{\nu \kappa}F_{\gamma \kappa}). \label{eq:8}
\end{align}

    \section{The leading higher derivative corrected KN geometry}
    
To set up our investigation, we first need to obtain the background geometry of the theory.Ref.~\cite{Cano:2019ore} established a method for deriving rotating black hole approximate solution at leading order in corrections beyond GR. The gravitational and Maxwell field solution of the 4-derivative extension of Einstein--Maxwell theory were subsequently derived in Ref.~\cite{Ma:2024ulp}. Following their approach, we re-evaluate the solution up to first order in $a$, $c_i$, and 14th order in the electric charge $Q$, as our aim is to study high charge effect on the gravito-electromagnetic perturbations of this slow rotation black hole. This requires $\chi \ll 1$ and $c_i \ll 1$ for all $i$. We now briefly review this metric solving procedure. Firstly,the higher-derivative geometry and Electromagnetic potential are assumed to take a form that preserves the event horizon structure of the KN black hole\eqref{eq:9}\eqref{eq:11}. Note that we adopt a metric ansatz which is of order $\mathcal{O(\chi)}$, and it's form differs slightly from those in Refs.~\cite{Cano:2019ore,Ma:2024ulp}; this choice of ansatz is crucial for avoiding spurious divergences in the effective potential of the perturbation equations,as well as for simplifying the calculation. 
    \begin{align}
ds^2 &= -\left(1+G_{1}\right)\left(1 - \frac{2Mr-Q^{2}}{r^2}\right)dt^2 - \left(1+G_{2}\right)\frac{2a\sin^2\theta(2Mr-Q^{2})}{r^2} dtd\phi + \left(1+G_{3}\right)r^2 \left(\frac{dr^2}{\Delta} + d\theta^2\right) \notag \\
&\quad +\left(1+G_{4}\right)r^2\sin^2\theta d\phi^2, \label{eq:9}
\end{align}
\begin{align}
 \quad \Delta = r^2 - 2Mr+Q^{2} . \label{eq:3.2}
\end{align}
\begin{align}\label{eq:11}
\bar{A_{\mu }} = -\frac{2Q}{r} \left \{  (1 + G_5), 0,0,- (1 + G_6)a\sin^2\theta\right\} ,
\end{align}
The metric correction functions \(G_i\) are found to always be expressible in the form of Eq.~\eqref{eq:12}.They are power series in \(\chi\) and \(\chi_Q\), and simultaneously polynomials in \(x\) and \(1/r\), with $G_i^{(n,p,j,k)}$ being the constant coefficients to be solved.
    \begin{align}\label{eq:12}
G_i = \sum_{n=0}^{\infty}G_i^{(n)}\chi^n,\quad i=1,2,3,4, 5,6
\end{align}

\begin{align}\label{eq:13}
\quad 
G_i^{(n)} = \sum_{p=0}^{n}\sum_{j=0}^{\infty }\sum_{k=0}^{k_{\mathrm{max}}}G_i^{(n,p,j,k)}x^p\chi _{Q}^{j}(\frac{M}{r})^{k},\quad x=\cos\theta
\end{align}
Here $\chi _{Q}=Q/M$ is the dimensionless charge parameter.We require \(\chi^2 + \chi_Q^2 \le 1\), with the equality corresponding to the extremal Kerr--Newman case.We truncate the series of $\chi$ to first order and $\chi_{Q}$ to 14th order.By inserting the ansatz \eqref{eq:9}\eqref{eq:11}\eqref{eq:12} into the linearized field equation ,the field equations are transformed to a series of algebraic equations, by solving them the value of the coefficients  $G_i^{(n,p,j,k)}$ can be obtained order by order in $(\chi ,x,\chi_Q,r)$.Readers can refer to \cite{Cano:2019ore,Ma:2024ulp} for more details.
 {\linespread{1}\selectfont
\begin{figure}[htbp]
    \centering
    \caption{Flowchart illustrating the perturbation procedure for the axial sector. The polar sector is treated in an analogous manner and is omitted for brevity.}
    \begin{tikzpicture}[node distance=1.5cm,
        every node/.style={align=center}, % 全局允许换行
        mybox/.style={draw, rounded corners, text width=4.5cm, align=center}
    ]
        
        % 中间的框
        \node (top) [mybox] 
            {1. Field equations of higher-derivative \eqref{eq:3}\eqref{eq:4}};
        
        % 左边的框：在 top 下方偏左
        \node (left1) [mybox, 
                      below left=1cm and -1cm of top] 
            {1.1. Einstein Maxwell part of field\\ equation, perturbated by \eqref{eq:14}\eqref{eq:15}};
            
        % 右边的框：在 top 下方偏右
        \node (right1) [mybox,
                       below right=1cm and -1cm of top] 
            {1.2. Higher derivative part of field\\ equation (coefficient of $\alpha$), perturbated by \eqref{eq:14}\eqref{eq:15}};
        
        \node (mid) [mybox, below right=2.5 cm and -1cm of left1]
            {2 The two dimensional perturbative\\ equations \eqref{eq:20}-\eqref{eq:25}, order $\epsilon$ (including \(\epsilon\alpha\))};
        
        \node (mid1) [mybox, below=1cm of mid]
            {3. Corrected coupled radial\\ equations \eqref{eq:26}-\eqref{eq:30} or \eqref{eq:31}\eqref{eq:32}};
        
        \node (mid2) [mybox, below=2.5 cm of mid1]
            {4. The two reduced corrected radial\\ equations};
        
        \node (left4) [mybox, below left=1cm and -1cm of mid2]
            {4.1. KN coupled radial\\ master equations Eq.(2) in \cite{Pani:2013wsa} };
        
        \node (right4) [mybox, below right=1cm and -1cm of mid2]
            {4.2. The corresponding correction\\ terms of KN equations};
        
        \node (right5) [mybox, below=2cm of right4]
            {4.3. The reduced correction\\ terms of equations};
        
        \node (right6) [mybox, below=7cm of mid2]
            {5. The corrected coupled\\ radial master equations \eqref{eq:33}};
        
        % 箭头连接（使用 rounded corners 让拐角变成小圆弧）
        \draw [->, rounded corners=4pt] (top.south)   |- (left1.east);
        \draw [->, rounded corners=4pt] (top.south)   |- (right1.west);
        
        \draw [->, rounded corners=4pt] (left1.south) -- 
            node[below, font=\footnotesize, align=center, pos=0.45] 
                {All quantities double expand\\ up to $\mathcal{O} (\epsilon \alpha )$} 
            (left1.south |- mid.north) -- (mid.north);
        
        \draw [->, rounded corners=4pt] (right1.south) -- 
            node[below, font=\footnotesize, align=center, pos=0.45] 
                {All quantities double expand\\ up to  $\mathcal{O} (\epsilon \alpha^{0} )$} 
            (right1.south |- mid.north) -- (mid.north);
        
        \draw [->] (mid.south) -- 
            node[right, font=\footnotesize, align=center] {Radial-angular\\ decouple} 
            (mid1.north);
        
        \draw [->] (mid1.south) -- 
            node[right, font=\footnotesize, align=center, text width=4cm] 
                {perturbatively around the route\\in \cite{Pani:2013wsa} of reducing the coupled equations, \\ treating the correction terms \\as functions independed of\\ $h_{0} ,h_{1} ,u$} 
            (mid2.north);
        
        \draw [->, rounded corners=4pt] (mid2.south) |- (left4.east);
        \draw [->, rounded corners=4pt] (mid2.south) |- (right4.west);
        
        \draw [->, rounded corners=4pt] (right4.south) -- 
            node[right, font=\footnotesize, align=center] {Rewrite $h_{0} ,h_{1} ,u$ as functions\\of $Z_{1}, Z_{2}$ with $\mathcal{O}(\alpha^{0})$\\ order relations between them} 
            (right5.north);
        
        \draw [->, rounded corners=4pt] (left4.south) |- 
            (right6.west);
        
        \draw [->, rounded corners=4pt] (right5.south) |-  node[pos=0.25, right, font=\footnotesize, align=center] {Deal high differential order\\ terms with KN master\\ equations of  $Z_{1}, Z_{2}$} (right6.east);
        
    \end{tikzpicture}\label{fig:1}
\end{figure}
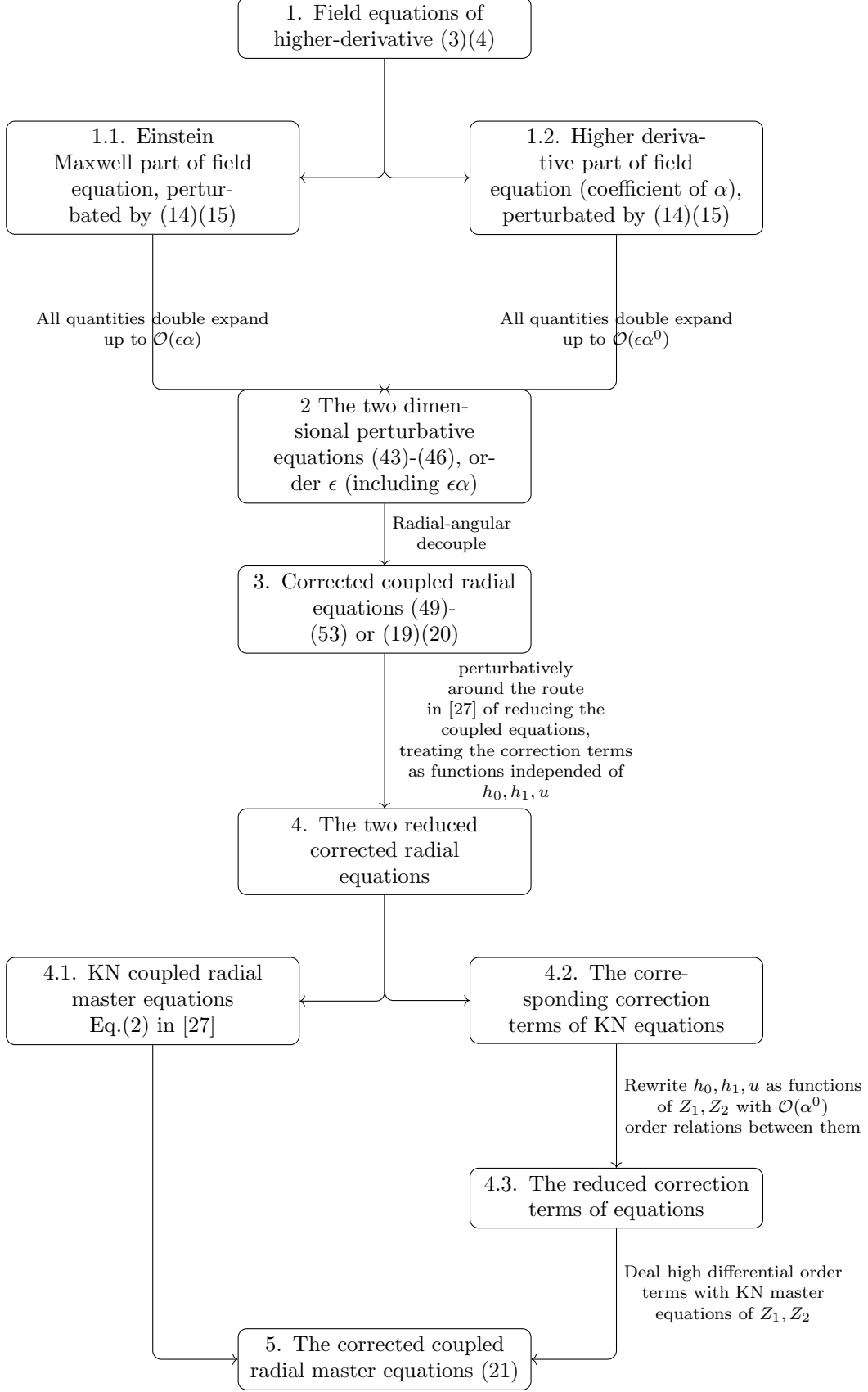}
\section{The perturbation scheme}
In this work,we make the first attempt to investigate the gravito-electromagnetic perturbations of this spacetime. An overview of the entire procedure --- from the initial perturbation setup to the final derivation of the coupled radial master equations --- is presented in the flowchart of Fig.~\ref{fig:1}, which serves as a roadmap for this paper. In our approach, the general metric and electromagnetic potential perturbations are decomposed into scalar, vector, and tensor harmonics. Working within the Regge--Wheeler gauge, we express the frequency \(\omega\) partial waves of the perturbations, in the following explicit form \cite{regge1957stability,Rosa:2011my,Pani:2013wsa}.
    \begin{equation}\label{eq:14}
\delta g_{\mu\nu} =\sum_{\ell,m}\begin{pmatrix}
H_0^{\ell m}(r)Y_{\ell m} & H_1^{\ell m}(r)Y_{\ell m} & h_0^{\ell m}(r)S_\theta^{\ell m} & h_0^{\ell m}(r)S_\phi^{\ell m} \\
* & H_2^{\ell m}(r)Y_{\ell m} & h_1^{\ell m}(r)S_\theta^{\ell m} & h_1^{\ell m}(r)S_\phi^{\ell m} \\
* & * & r^2 K^{\ell m}(r)S_{\theta \theta}^{\ell m} & 0 \\
* & * & * & r^2 K^{\ell m}(r)S_{\phi \phi}^{\ell m}
\end{pmatrix}
e^{-i\omega t}
 \end{equation}
 \begin{equation}\label{eq:15}
 \delta A_{\mu} = e^{-i\omega t} \sum_{\ell,m} 
\begin{pmatrix}

0 ,
0 ,
\dfrac{u^{\ell m}(r)}{\Lambda} S_{\theta}^{\ell m},
\dfrac{u^{\ell m}(r)}{\Lambda} S_{\phi}^{\ell m}\end{pmatrix}
 + \begin{pmatrix}
\dfrac{j^{\ell m}(r)}{r} Y_{\ell m} ,
\dfrac{k^{\ell m}(r)}{r F(r)} Y_{\ell m} ,
\dfrac{y^{\ell m}(r) }{\Lambda}Y_{,\theta}^{\ell m} ,
\dfrac{y^{\ell m}(r) }{\Lambda}Y_{,\phi}^{\ell m}
\end{pmatrix}
\end{equation}
 with
 \begin{align}\label{eq:16}
S_a^{\ell m} &= {\epsilon_a} ^b\partial_ b Y^{\ell m},\text{vector}\quad(-1)^{\ell+1} \quad \quad {Y^{\ell m}}_{,a} = \partial_ a Y^{\ell m},\text{vector}\quad(-1)^\ell \nonumber\\
S_{ab}^{\ell m} &= \gamma _{ab} Y^{\ell m},\text{tensor}\quad(-1)^\ell \quad \quad\quad\quad Y^{\ell m} ,\text{scalar}\quad(-1)^\ell \qquad a,b=\theta,\phi
\end{align}
where \(\gamma_{ab}\) is the induced metric and \(\epsilon_{ab}\) is the mixed-index Levi-Civita tensor on the unit two-sphere \(S^2\), with \(\Lambda = \ell(\ell+1)\), and \(Y_{\ell m}\) being the scalar spherical harmonics. The symbol \(*\) denotes the corresponding quantity determined by metric symmetry. Each harmonic in Eq.\ref{eq:16} carries a natural spin weight related to its transformation behavior under rotations of the sphere, as well as a definite parity: odd (axial) modes transform as \(f \to (-1)^{\ell+1} f\), while even (polar) modes transform as \(f \to (-1)^{\ell} f\), under the mapping \((\theta,\phi) \to (\pi-\theta, \phi+\pi)\). For notational brevity,in the remaining part of this paper, we omit the summation operator \(\sum_{\ell,m} e^{-i\omega t}\) and the superscripts \(\ell, m\) on the perturbation functions, except that we write \(Y^{\ell}\) in place of \(Y_{\ell m}\),which is justified by the fact that, owing to the axisymmetry of the background spacetime, modes with different \(m\) decouple from one another.We expand the full metric and electromagnetic potential as
    \begin{equation}\label{eq:18}
    \centering
	g_{\mu \nu } =\bar{g} _{\mu \nu}+\epsilon \delta g_{\mu \nu }\quad\quad\quad\quad
    \bar{g} _{\mu \nu}=g_{\mu \nu}^{(0)} + \alpha g_{\mu \nu}^{(1)}
    \end{equation}
     \begin{equation}\label{eq:19}
	A_{\mu } =\bar{A_{\mu }} +\epsilon \delta A_{\mu }\quad\quad\quad\quad
    \bar{A_{\mu }} =A_{\mu }^{(0)} + \alpha A_{\mu }^{(1)}
    \end{equation}
Here, \(\bar{g}_{\mu\nu}\) and \(\bar{A}_{\mu}\) denote the corrected background metric and electromagnetic potential, respectively, while \(g_{\mu\nu}^{(0)}\) and \(A_{\mu}^{(0)}\) represent their KN counterparts. The quantities \(g_{\mu\nu}^{(1)}\) and \(A_{\mu}^{(1)}\) account for the deviations from the KN solution induced by the higher-derivative corrections, which can be readily obtained from Eqs.~\eqref{eq:9} and~\eqref{eq:11}. and \(\delta g_{\mu\nu}\) and \(\delta A_{\mu}\) are the metric and electromagnetic perturbations about the corrected background.Here,we introduce the bookkeeping parameters \(\alpha\) and \(\epsilon\) to tag the higher-derivative corrections and the perturbations, respectively; these parameters are eventually set to unity. Substituting the above decompositions \eqref{eq:18}\eqref{eq:19} into the field equations ~\eqref{eq:3} and~\eqref{eq:4}, and expanding all relevant quantities --- including the Christoffel symbols, the Ricci tensor, and the entire field equations --- consistently to order \(\mathcal{O}(\epsilon\alpha)\), we obtain the corresponding perturbed geometric quantities, which are of order \(\epsilon\) (including order \(\epsilon\alpha\)).The field equations at
order \(\epsilon\) depict the dynamics of the metric perturbations.

During this process,for the Einstein--Maxwell part of the perturbed field equations, the zeroth-order terms in the higher-derivative parameters \(c_i\) coincide with those of the standard KN BH,and the first-order contributions arise from the corrections to the metric and electromagnetic potential.While
for the higher-derivative part of the field equations, it suffices to evaluate the background quantities using the pure KN solution, since these terms are intrinsically of order \(\alpha\). Inserting the full corrected geometry would introduce differences that enter only at \(\mathcal{O}(\alpha^2)\), which are beyond the linear order in \(\alpha\) considered in this work.

\section{Derive the coupled radial master equations}
The above perturbation procedure derives 10 gravitational and 4 electromagnetic equations. Through a radial-angular decoupling procedure (see Appendix A.1 for details), we obtain the purely radial coupled equations Eqs.~\eqref{eq:26}--\eqref{eq:30}.In this section, we describe the procedure for reducing these equations with a fix parity to corrected coupled radial master equations. We focus primarily on the axial (odd-parity) sector, as the treatment of the polar (even-parity) sector follows essentially the same strategy but is considerably more involved and time-consuming. Each of Eqs.~\eqref{eq:26}--\eqref{eq:30} can be decomposed into a zeroth-order part, which coincides with the corresponding KN equation, and a first-order correction term arising from the higher-derivative couplings. To render the reduction procedure transparent, we rewrite the equations in the following form.
\begin{equation}\label{eq:31}
\mathcal{D}^{(0)}_{\mathcal{L}} (h_{0} ,h_{1} ,u )+\alpha \mathcal{D}^{(1)}_{\mathcal{L}} (h_{0} ,h_{1} ,u )=0\quad\quad \mathcal{L}=1,2,3,4
\end{equation}

\begin{equation}\label{eq:32}
\mathcal{D}^{(0)}_{\mathcal{I}} (H_{0} ,H_{1} ,H_{2} ,K ,j,k ,y  )+\alpha \mathcal{D}^{(1)}_{\mathcal{I}} (H_{0} ,H_{1} ,H_{2} ,K ,j,k ,y   )=0\quad\quad \mathcal{I}=1,2...10
\end{equation}

Here, \(\mathcal{D}^{(0)}_{L}\) and \(\mathcal{D}^{(1)}_{L}\) represent the operators in KN equations and the first-order corrections, respectively.The reduction proceeds in two stages. In the first stage, we linearly perturbative around the route in Ref\cite{Pani:2013wsa} of reducing these equations to two coupled master equations.For the zeroth order part,\(h_0, h_1, u_4\) are finally reduced to \(Z_1\) and \(Z_2\). But the correction terms are treated as functions independent of \(h_0, h_1, u_4\), we retain their form throughout this stage.Upon finishing the perturbative reduction around the KN background stage, we derive the corrected two coupled equations, whose zeroth-order part exactly matches Eq.(2) of Ref.~\cite{Pani:2013wsa}. Then in the second stage, we deal with the higher-derivative correction terms. We express \(h_0, h_1, u_4\) in terms of \(Z_1, Z_2\) via the zeroth-order relations, and use the zeroth-order equations for \(Z_1, Z_2\) to reduce the derivative order. This yields the corrected coupled radial master equations~\eqref{eq:33}, which reduce to the KN equations [Eq.(2) of Ref.~\cite{Pani:2013wsa}] when the higher-derivative parameters vanish. The resulting equations for both polar and axial parities are among the main results of this work; their explicit forms are provided in the Supplemental Material due to their length.

\begin{equation}\label{eq:33}
\boldsymbol{\mathcal{D}}^{(0)\pm } (\boldsymbol{Z}^{\pm })+\alpha(\boldsymbol{P}^{\pm } \boldsymbol{Z}^{'\pm }+\boldsymbol{Q}^{\pm } \boldsymbol{Z}^{\pm }) =0
\end{equation}
with
\[
\boldsymbol{Z}=
\begin{pmatrix} Z _1 \\ Z _2 \end{pmatrix},\quad
\boldsymbol{P} = \begin{pmatrix} P_{11} & P_{12} \\ P_{21} & P_{22} \end{pmatrix},\quad
\boldsymbol{Q} = \begin{pmatrix} Q_{11} & Q_{12} \\ Q_{21} & Q_{22} \end{pmatrix}.
\]
Here, the superscripts \(+\) and \(-\) denote the even- and odd-parity sectors, respectively. The matrix operator \(\boldsymbol{\mathcal{D}}^{(0)}\) represents the operator acting on \(\boldsymbol{Z}\) in the KN equations, which can be readily derived from Eq. (2) of Ref.~\cite{Pani:2013wsa}. The second and third terms are the higher-derivative correction contributions, with all components of the matrices \(\boldsymbol{P^\pm}\) and \(\boldsymbol{Q^\pm}\) being functions of \(r\).Owing to the complexity of these equations, and since we are concerned only with linear-order higher-derivative effects, we perform the above reduction procedure separately for each \(c_i\). The complete equations can then obtained by simply summing the contributions from all \(c_i\) corrections. This linear superposition rule also applies to the frequency shifts discussed below, as they are linear in \(c_i\) as well. We find that the parameter \(c_1\) does not correct either the KN geometry or the final radial perturbation equations for both parities.Consequently, the QNFs remain unmodified by $c_1$.

\section{Matrix-valued asymptotic iteration method }

Considering the lengthy coupled equations and seven parameters in our model to investigate, conventional techniques for computing QNMs---such as the pseudo-spectral method \cite{Jansen:2017oag}, the WKB approach \cite{SchutzWill1985,IyerWill1987}, and the matrix-valued continued-fraction method \cite{Leaver1985,Pani:2013pma,Pani:2013wsa}---appear either impractical to implement or prohibitively time-consuming. We therefore adopt the asymptotic iteration method (AIM), which we extend to handle the coupled second-order ordinary differential equations governing our system. To validate our results, we also perform cross-checks using direct integration. While the original AIM was formulated for single ODEs \cite{ciftci2005construction} (see reiview \cite{Cho:2011sf}) and later extended to directly solve the coupled first-order Dirac equation \cite{Hernandez-Velazquez:2021zoh}.The AIM for single ODE have been widely used to search for QNFs\cite{barakat2006asymptotic,Cho:2009cj,Mamani2022,Shi2025},the generalization to coupled second-order ODEs (the following form) has not, to our knowledge, been previously reported. This constitutes the first application of AIM to such systems in the context of black-hole QNMs.
\begin{align}
\Psi _1''(x) &+ W_1(x)\Psi _1'(x) + R_1(x)\Psi _1(x) + W_2(x)\Psi _2'(x) + R_2(x)\Psi _2(x) = 0,
\\
\Psi _2''(x) &+ J_1(x)\Psi _1'(x) + K_1(x)\Psi _1(x) + J_2(x)\Psi _2'(x) + K_2(x)\Psi _2(x) = 0,
\end{align}

where the coefficient functions also depend on the quasinormal mode frequency \(\omega\). Applying the same manipulation to the perturbation equations, we rewrite the two equations in a compact matrix form.
\begin{equation}
\boldsymbol{\Psi }''(x) = \boldsymbol{A}_0(x)\boldsymbol{\Psi }'(x) + \boldsymbol{B}_0(x)\boldsymbol{\Psi }(x),
\end{equation}
with
\[
\boldsymbol{\Psi }=
\begin{pmatrix} \Psi _1 \\ \Psi _2 \end{pmatrix},\quad
\boldsymbol{A}_0 = -\begin{pmatrix} W_1 & W_2 \\ J_1 & J_2 \end{pmatrix},\quad
\boldsymbol{B}_0 = -\begin{pmatrix} R_1 & R_2 \\ K_1 & K_2 \end{pmatrix}.
\]
First we need to factor out the boundary asymptotic behavior of the wave functions by means of the Frobenius analysis and impose the boundary condition, We then map the coordinate \(x\) onto the unit interval \(u \in (0,1)\), which yields a regularized version of the wave equation that is smooth on the entire domain.
\begin{equation}\label{eq:37}
\boldsymbol{\psi}''(u) = \tilde{\boldsymbol{A}}_0(u)\boldsymbol{\psi}'(u) + \tilde{\boldsymbol{B}}_0(u)\boldsymbol{\psi}(u),
\end{equation}
Differentiate this equation and substitute \eqref{eq:37} to eliminate the second differential order terms,then repeat this process for $n-1$ times,One can obtains 

\begin{equation}
\boldsymbol{\psi}^{(n+2)}(u) = \tilde{\boldsymbol{A}}_n(u)\boldsymbol{\psi}'(u) + \tilde{\boldsymbol{B}}_n(u)\boldsymbol{\psi}(u),
\end{equation}
with the matrix coefficients recursion relations
\begin{align}\label{eq:39}
\tilde{\boldsymbol{A}}_n &= \tilde{\boldsymbol{A}}_{n-1}' + \tilde{\boldsymbol{B}}_{n-1} 
+  \tilde{\boldsymbol{A}}_{n-1}\tilde{\boldsymbol{A}}_0,
\end{align}
\begin{align}\label{eq:40}
\tilde{\boldsymbol{B}}_n &= \tilde{\boldsymbol{B}}_{n-1}' + \tilde{\boldsymbol{A}}_{n-1}\tilde{\boldsymbol{B}}_0 ,
\end{align}
Imposing the quantilization condition---namely, that when the iteration number $n\to \infty$, the solutions become linearly dependent, which means the following matrix termination equation
\begin{equation}\label{eq:42}
\tilde{\boldsymbol{A}}_{n-1}\tilde{\boldsymbol{B}}_n  
- \tilde{\boldsymbol{A}}_n\tilde{\boldsymbol{B}}_{n-1}= \boldsymbol{0}.
\end{equation}
this approximately equals the following numerically solvable equation
\begin{equation}
\det\left(
\tilde{\boldsymbol{A}}_{n-1}\tilde{\boldsymbol{B}}_n  
- \tilde{\boldsymbol{A}}_n\tilde{\boldsymbol{B}}_{n-1} 
\right) = 0,
\end{equation}
which is the determinant of \eqref{eq:42}.
The root is exactly the eigenvalue of the coupled ODEs. This method requires the equation \eqref{eq:37} to be sufficiently smooth and differentiable at each iteration step; otherwise, numerical instability may arise. This is why we fact out the boundary asymptotic behaviour of the wave functions at the outset. Following the improved AIM approach \cite{Cho:2009cj}, we expand the matrix coefficient functions as Taylor series at an appropriate point $\xi$, preferably chosen far from the boundaries.

\begin{align}\label{eq:43}
\tilde{\boldsymbol{A}}_n(u) &= \sum_{k=0}^{\infty} \boldsymbol{c}_n^k \,(u-\xi)^k
\end{align}
\begin{align}\label{eq:44}
\tilde{\boldsymbol{B}}_n(u) &= \sum_{k=0}^{\infty} \boldsymbol{d}_n^k \,(u-\xi)^k,
\end{align}
Here, \(\boldsymbol{c}_n^i\) and \(\boldsymbol{d}_n^i\) denote the coefficient matrices of the Taylor expansion. Inserting the matrix-valued series into Eqs.~\eqref{eq:39} and~\eqref{eq:40}, by matching powers of \(u\),the following constant matrix recursion relations can be yield:
\begin{align}
\boldsymbol{c}_n^k &= (k+1)\boldsymbol{c}_{n-1}^{k+1} + \boldsymbol{d}_{n-1}^k 
+ \sum_{i=0}^{k} \boldsymbol{c}_{n-1}^{k-i}\boldsymbol{c}_0^i,
\\
\boldsymbol{d}_n^k &= (k+1)\boldsymbol{d}_{n-1}^{k+1} 
+ \sum_{i=0}^{k} \boldsymbol{c}_{n-1}^{k-i}\boldsymbol{d}_0^i.
\end{align}
Employing this numerical iteration instead of analytical derivative iteration greatly simplifies the calculation and improves numerical efficiency. Moreover, the quantization condition is reduced to a simple algebraic equation truncating at a large $n=n_{max}$

\begin{equation}
\det\left(
\boldsymbol{c}_{n-1}^0\boldsymbol{d}_n^0  
-\boldsymbol{c}_n^0 \boldsymbol{d}_{n-1}^0 
\right) = 0.
\end{equation}
 The eigenvalues of the matrix are exactly the QNFs of the coupled perturbation system. It should be noted that \(n_{\max}\) must not exceed the expansion order in Eqs.~\eqref{eq:43} and~\eqref{eq:44}.As noted in Ref.~\cite{Cho:2009cj}, a higher expansion order does not necessarily guarantee better convergence. For our coupled system, we find that the optimal expansion order is below 24.

\section{QNFs results}

 The functions \(Z_1\) and \(Z_2\) in Eq.\eqref{eq:33} are both Grav-EM coupled and are predominantly associated with the Grav and EM fields, respectively.Consequently, we obtain two QNFs for each parity of Eq.\eqref{eq:33}, which are predominantly associated with the Grav mode and EM mode, respectively.Returning to the motivation of this work, we now examine the effects of the higher-derivative corrections on each mode.
 \subsection{Isospectral}
\begin{figure}
		\includegraphics[width=0.5\linewidth]{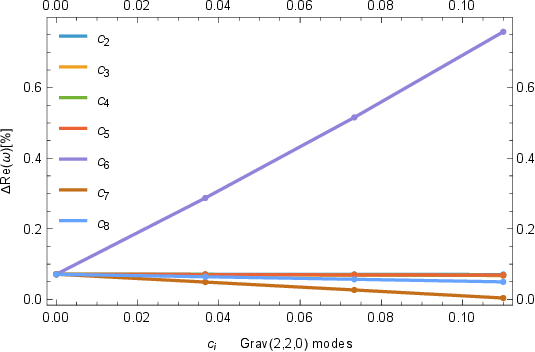}
	\includegraphics[width=0.5\linewidth]{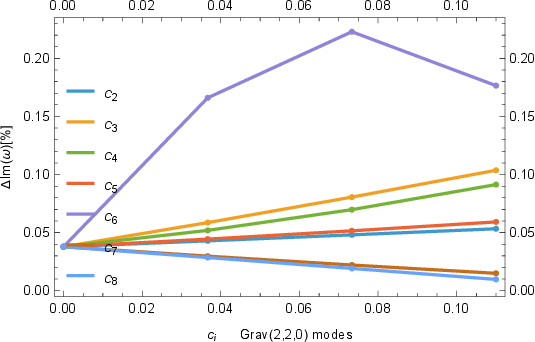}
    \vspace{0.5cm}
		\includegraphics[width=0.5\linewidth]{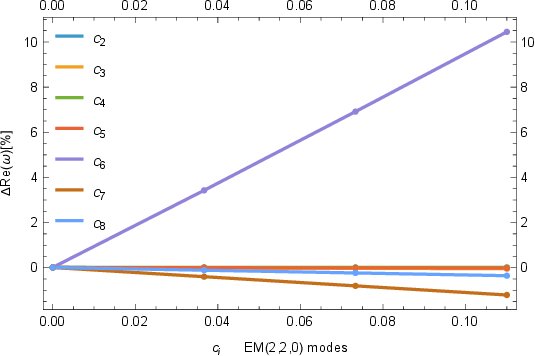}
	\includegraphics[width=0.5\linewidth]{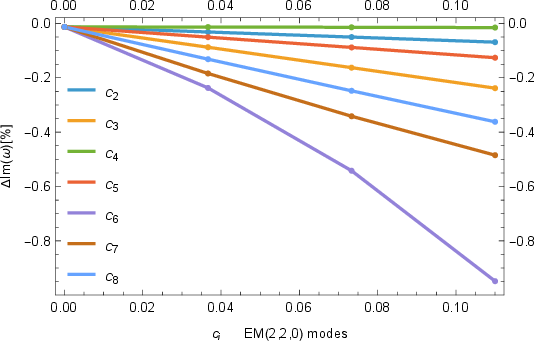}
     \vspace{0.5cm}
	\includegraphics[width=0.5\linewidth]{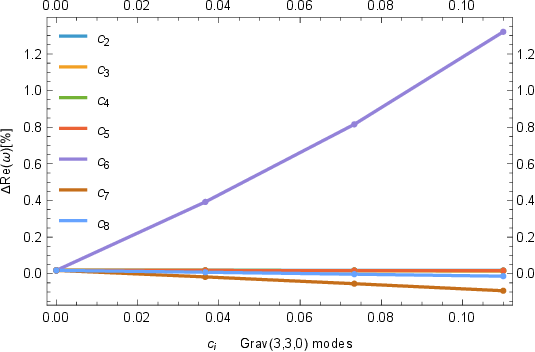}
		\includegraphics[width=0.5\linewidth]{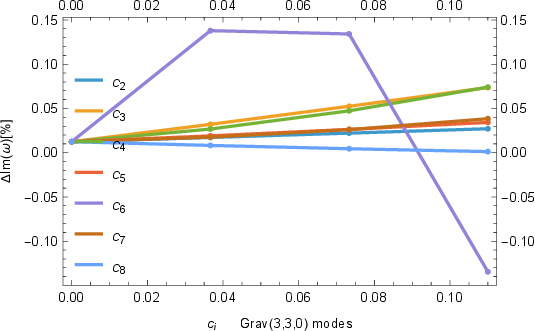}
         \vspace{0.5cm}
	\includegraphics[width=0.5\linewidth]{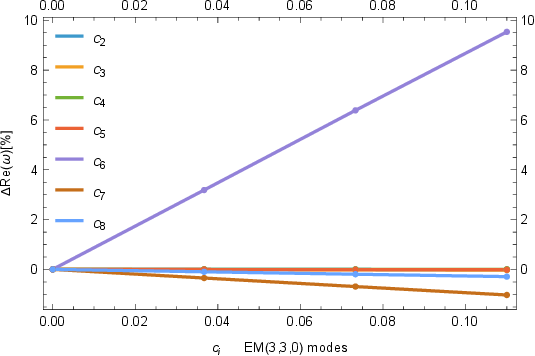}
	\includegraphics[width=0.5\linewidth]{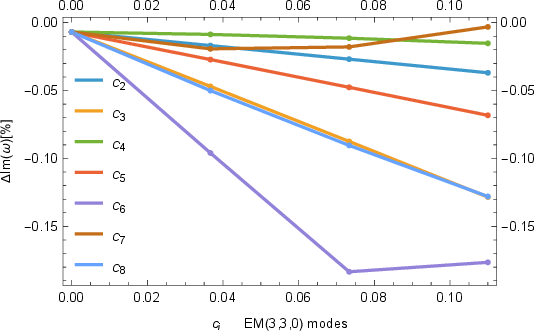}
    \\
	\caption{ISO-breaking percentage given by Eq.\eqref{eq:48} with respect to higher derivative parameters $c_i$,the left panels display the real part and the right panels for the imaginary part.We use $\chi=11/150,\chi_Q=0.31$} 
	\label{fig:2}
\end{figure}

\begin{figure}
		\includegraphics[width=0.5\linewidth]{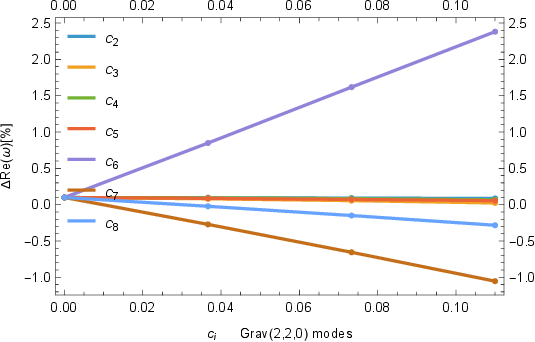}
	\includegraphics[width=0.5\linewidth]{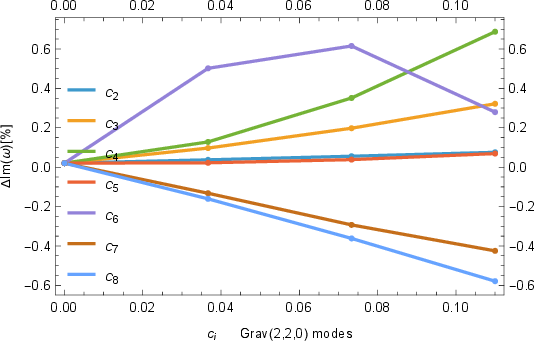}
    \vspace{0.5cm}
		\includegraphics[width=0.5\linewidth]{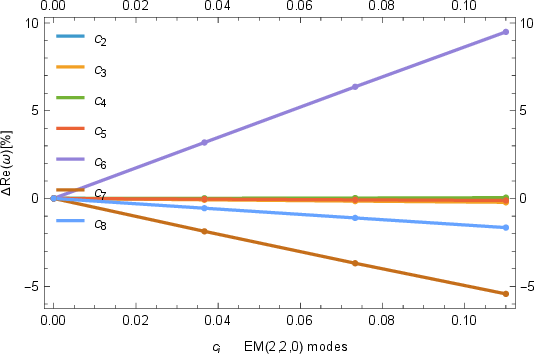}
	\includegraphics[width=0.5\linewidth]{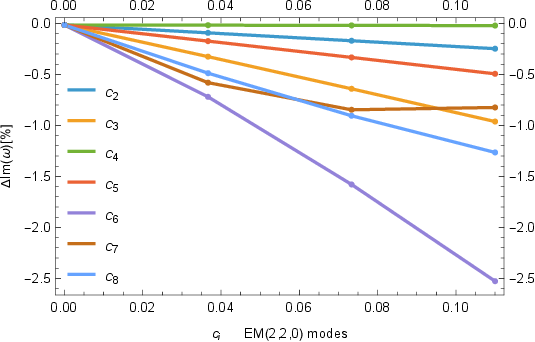}
     \vspace{0.5cm}
	\includegraphics[width=0.5\linewidth]{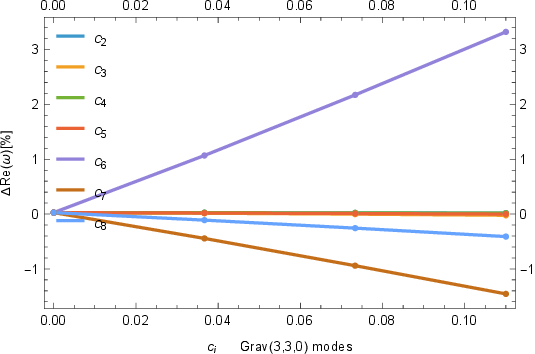}
		\includegraphics[width=0.5\linewidth]{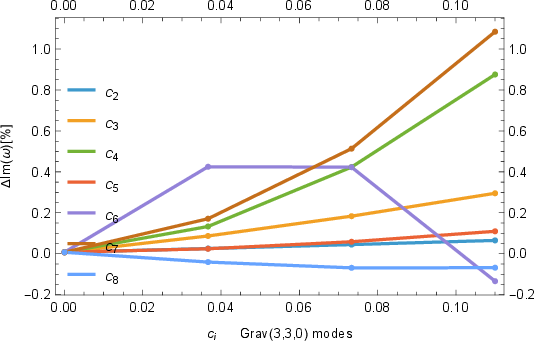}
         \vspace{0.5cm}
	\includegraphics[width=0.5\linewidth]{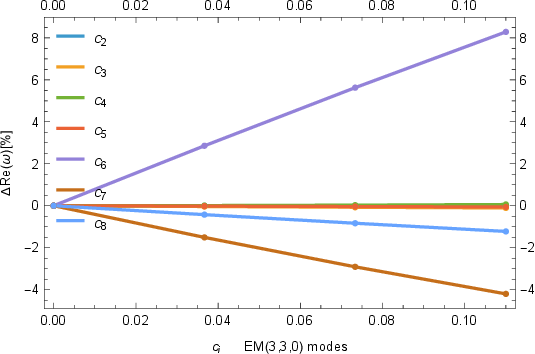}
	\includegraphics[width=0.5\linewidth]{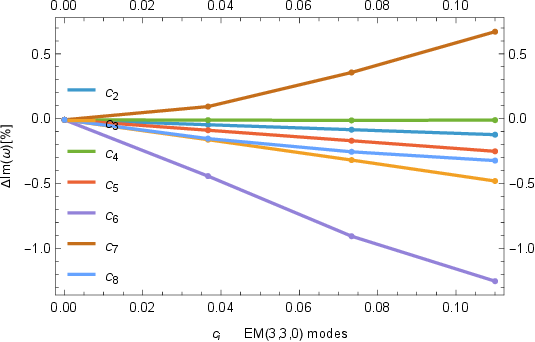}
    \\
	\caption{ISO-breaking percentage  given by Eq.\eqref{eq:48} with respect to higher derivative parameters $c_i$,the left panels display the real part and the right panels for the imaginary part.We use $\chi=11/150,\chi_Q=0.61$} 
	\label{fig:3}
\end{figure}

  For the Kerr--Newman black hole, isospectrality has been established up to first order in \(\chi\) \cite{Pani:2013wsa}. A natural question then arises: can the higher-derivative corrections induce a splitting of the QNF degeneracy between the two parity sectors? To address this, we display the differences between the fundamental QNFs of the two parities versus $c_i\in(0,0.11)$ for \(l=m=2,3\) modes with \(\chi = 11/150\) in Figs.~\ref{fig:2} and~\ref{fig:3}, where \(\chi_Q\) is set to \(0.31\) and \(0.61\), respectively. The percentage of ISO breaking is defined as follows:

\begin{equation}\label{eq:48}
 \Delta \Pi_k(\omega)=\frac{\Pi_k(\omega_{\rm polar})-\Pi_k(\omega_{\rm axial})}{\Pi_k(\omega_{\rm axial})}\times 100\%,\quad k=R,I
\end{equation}
Here, \(\omega_{\rm axial}\) and \(\omega_{\rm polar}\) denote the axial and polar QNFs, respectively, and \(\Pi_R\) and \(\Pi_I\) are operators that extract the real and imaginary parts. From  Figs.~\ref{fig:2} and~\ref{fig:3}, we observe that :as in Ref.~\cite{Brito:2018hjh},the splitting of the EM modes is typically significantly larger than that of the Grav modes; the electric charge generally enhances the difference between the QNFs of the two parity sectors;the higher-derivative parameter \(c_6\), which corresponds to the term \(R_{\mu\nu\gamma\kappa}F^{\mu\nu}F^{\gamma\kappa}\) in the action, leads to a most pronounced splitting across all modes, especially in the real part, where \(\Delta \operatorname{Re}(\omega)\) can reach up to \(10\%\) for the EM modes,It is worth mentioning that Ref.~\cite{Boyce:2025fpr} derived an analytical expansion for the frequency shifts of the zero-damping modes in near-extremal EFT-corrected Reissner--Nordstr\"om black holes. Their result indicates that the \(c_7\) and \(c_8\) terms contribute at an order larger than the other parameters, including \(c_6\). This may suggest that the enhanced near-horizon symmetry of the near-extremal geometry suppresses the relative spectral splitting induced by \(c_6\);In our results for \(\chi_Q = 0.61\), the EM modes with non-linear electromagnetic coupling parameters \(c_7 = 0.11\) or \(c_8 = 0.11\), as well as the Grav modes with \(c_7 = 0.11\), also exhibit \(\Delta \operatorname{Re}(\omega)\) values that are significantly larger than the associated relative errors. Our results for \(c_7\) are consistent with those obtained in a spherically symmetric system studied in Ref.~\cite{Nomura:2021efi}. For the remaining parameter combinations, the ISO breaking is below \(1\%\) for either the real or imaginary parts, and there does not seem to exist a clear separation between these values and the numerical uncertainties. Therefore, in the next section, for the modes that exhibit significant splitting, we compute the fundamental QNFs for both parities with \(l=m=2,3\), while for the remaining cases we present only the axial QNFs.

\subsection{QNFs shifts}
Since the numerical QNFs are available only at discrete parameter values, we provide an analytical interpolation via polynomial fitting to enhance the generality of our results. We expand the real and imaginary parts of the QNFs as
\begin{equation}\label{eq:49}
  \log\left|M\omega_{R,I}\right| = \sum_{n=0}^{1} \sum_{j=0}^{1} \sum_{k=0}^{5} d^{(n,j,k)}_{R,I} \, c_i^{\,n} \chi^{\,j} T_k(2\chi_Q - 1), \qquad i = 2,3\ldots8,
\end{equation}
where \(T_k\) are the Chebyshev polynomials of the first kind. We display the fitting results for the coefficients $d^{(n,j,k)}_{R,I}$ in Tabs.~\ref{tab:1}--\ref{tab:8} of appendix. We then compare the QNFs obtained from the fitting formula---via interpolation at $\chi_Q = 0.08, 0.28, 0.48$ and extrapolation at $\chi_Q = 0.68$---with those computed by the numerical method---direct integration (DI) \cite{Pani:2013pma,Pani:2013wsa} for the axial sector and our AIM implementation for the polar sector. The relative errors and the percentage shifts of the QNFs are evaluated using the following two formulae, respectively.

\begin{equation}
\Delta_{\rm{ER}}^{(k)}=\left|\frac{\Pi_k(\omega _{\rm{fit}})-\Pi_k(\omega _{\rm{Num}})}{\Pi_k(\omega _{\rm{Num}})}\right| \times 100\%,\quad k=R,I 
 \end{equation}
 \begin{equation}
\Delta_{\omega}^{(k)}=\frac{ \Pi_k(\omega_{\rm{fit}}^{\rm{EFT}})-\Pi_k(\omega_{\rm{fit}}^{\rm{KN}})}{\Pi_k(\omega_{\rm{fit}}^{\rm{KN}})}\times 100\% ,\quad k=R,I
\end{equation}
Here, $\omega_{\rm{fit}}^{\rm{KN}}$ and $\omega_{\rm{fit}}^{\rm{EFT}}$ denote the Kerr--Newman frequency and the corrected frequency in the effective-field-theory framework, respectively, both obtained from the fitting formula in Eq.~\eqref{eq:49}. The comparison results for $c_i$ ($i=2,3,\dots,8$) with $\chi = 0.08$ and $c_i = 0.09$ are presented in Tabs.~\ref{tab:9}--\ref{tab:16} of the appendix. As mentioned above,a large black-hole charge $Q$ significantly enhances the QNF shifts relative to the KN values. For $\chi_Q = 0.48$ and $0.68$, the relative errors are mostly smaller than one-tenth of the frequency-shift percentages. Given the projected sensitivity of future gravitational-wave detectors, this may offer a new avenue for probing higher-derivative effects through gravitational-wave observations.
Now let us examine the invariants under field redefinitions in effective field theory \cite{Ma:2024ulp},by imposing the following transformation of the metric:
\begin{equation}
g_{\mu\nu} \rightarrow g_{\mu\nu} + \delta_1 R_{\mu\nu} + \delta_2 g_{\mu\nu} R+ \delta_3 F_{\mu\rho} F_{\nu}^{\rho} + \delta_4  g_{\mu\nu}F^2
\end{equation}
which leads to a shift of the higher derivative parameters in Eq.\ref{eq:1}
\begin{align}
c_1 \rightarrow c_1 + \frac{1}{2}\delta_1 + \delta_2, \qquad 
c_2 \rightarrow c_2 - \delta_1, \qquad 
c_3 \rightarrow c_3, \qquad
c_4 \rightarrow c_4 - \frac{1}{8}\delta_1 + \frac{1}{2}\delta_3 + \delta_4, \nonumber \\
c_5 \rightarrow c_5 + \frac{1}{2}\delta_1 - \delta_3, \qquad
c_6 \rightarrow c_6, \qquad
c_7 \rightarrow c_7 - \frac{1}{8}\delta_3, \qquad
c_8 \rightarrow c_8 + \frac{1}{2}\delta_3,
\end{align}
It is easy to show that the following combination is invariant under the transformation given above, and hence qualifies as an observable physical quantity.
\begin{align}\label{eq:42}
\alpha_1 = c_3 \nonumber\\
\alpha_2 = c_6 \nonumber\\
\alpha_3 = c_2 + 4c_3 + 2c_5 + 2c_6 + 16c_7 + 8c_8 \nonumber \\
\alpha_4 = c_2 + 2c_5 + 4c_8
\end{align}
We observe that the above equation is irrelevant to \(c_4\). Moreover, the expressions for \(\alpha_3\) and \(\alpha_4\) indicate that the effect of \(c_2\) is equivalent to that of \(2c_5\). Regarding the former, our results in Tab.~\ref{tab:9}--\ref{tab:16} show that the QNF shifts induced by \(c_4\) are much smaller (by about an order of magnitude) than those induced by the remaining parameters, which supports the invariance under field redefinitions. For the latter, our results do not provide sufficient evidence, though this does not constitute counterevidence.
We also find that, in all the ISO-breaking cases discussed above, the frequency shifts of the two parity sectors occur in opposite directions when $\chi_Q \ge0.28 $, as shown in Tab.~\ref{tab:9}--\ref{tab:16} of the appendix. This provides a direct interpretation of the frequency splitting. This finding is consistent with the result in Ref.~\cite{Cano:2024ezp}.
\section{Conclusion and discussion}
We have investigated linear Gravito-Electromagnetic perturbations of a slowly rotating, electrically charged black hole within the general EFT extensions of GR---the leading 4-derivative corrected Einstein--Maxwell system.After performing the radial-angular decoupling and a cumbersome reduction procedure, we obtain the coupled radial master perturbation equations for both axial and polar sectors. By extending the asymptotic iteration method to handle coupled second-order ODEs, we are able to efficiently search for QNFs of this system. Our results reveal isospectrality breaking induced by the higher-derivative parameters $c_6$, $c_7$, and $c_8$, particularly for EM modes of high charge BH. Here, \(c_7\) and \(c_8\) correspond to pure nonlinear electromagnetic terms in the action, whereas \(c_6\) couples the Riemann tensor to the Maxwell field strength. Notably,the largest isospectrality-breaking percentage induced by \(c_6 \) can reach up to $10\%$ for the real parts of the electromagnetic modes with \(c_6 = 0.11\).The different responses of the two parity to higher derivative corrections implies that the above three parameters introduce parity-dependent terms into the perturbation equations. Consequently, the equations for the two parity sectors no longer possess the transformation symmetry that relates them in the KN case. In addition, we obtain analytic expressions for the QNFs via a high-accuracy polynomial fitting.The relative errors are mostly smaller than $1/10$ of QNFs shifts percentage, which induced by higher derivative parameters for high charge $Q$.Given the unprecedented sensitivity of next-generation gravitational-wave detectors, such deviations and ISO-breaking phenomenon could serve as a potential avenue for testing modified gravity theories using ringdown signals.We also find: In the ISO-splitting cases, the QNF shifts from GR take opposite directions for \(\chi_Q \gtrsim 0.28\), offering a direct interpretation of the degeneracy splitting. This is consistent with Ref.~\cite{Cano:2024ezp}; Moreover, the \(c_4\)-induced shifts are about an order of magnitude smaller than those from the other parameters, supporting the view that \(c_4\) corresponds to a redundant degree of freedom in the EFT.We hope that our work will facilitate future investigations of QNMs in modified gravity theories.

Nevertheless,there remain promising directions for future work. For instance, extending the perturbation analysis to higher orders in angular momentum would allow us to examine whether the $c_2$-$c_5$ parameters induce significant isospectrality breaking at higher order.However,this appears prohibitively cumbersome within a purely analytical framework. A complete numerical approach would likely be a suitable strategy to address this challenge,and simultaneously, to provide an independent verification of our current perturbative results.An analytical result presented in Ref.~\cite{Boyce:2025fpr} shows that, for the 4-derivative-corrected near-horizon, near-extremal Reissner--Nordstr\"om black hole, the ISO-breaking of the zero-damping modes induced by \(c_7\) and \(c_8\) provides the leading contribution. This may indicate that the enhanced symmetry of the near-horizon, near-extremal geometry suppresses the more significant ISO-breaking that would otherwise be induced by \(c_6\). We leave a detailed verification of this possibility to future work.
\section*{Acknowledgements}
We are grateful to Liang Ma for helpful discussion.This research was supported by the National Natural Science Foundation of China (Grant No. 12265007) and the Guizhou Provincial Major Scientiﬁc and Technological Program (Grant
No. XKBF (2025)010).
\appendix
% 此命令会让后续 \section 编号变为 A, B, C...

\section{Appendix} 
% 自动编号为 "A"，自动加入目录

\subsection{Decoupling of the radial and angular equations}
After the perturbation procedure in Sect.4, we obtain a system of 14 coupled perturbation equations in the radial and angular dimension: 10 from the gravitational and 4 from the Maxwell sector.Following the procedure outlined in Refs.~\cite{Pani:2013wsa,Srivastava:2021imr}, we then eliminate the angular dependence to derive the pure radial equations. The 14 equations can be classified into three groups according to their structure, as follows:

Group I:consists of six equations --- namely, the \((tt)\), \((tr)\), \((rr)\), and \((\theta\theta) + (\phi\phi)/\!\sin^2\theta\) components of the gravitational equation, together with the \(t\) and \(r\) components of the Maxwell equation. These are collectively denoted as \(\delta E^{I}\) with \(I = 1, \dots, 6\), and are presented explicitly in Eq.~\eqref{eq:20}.
\begin{equation}
\delta E^{(I)} \equiv (A^{(I)}_{\ell} + \tilde{A}^{(I)}_{\ell} \cos \theta) Y^{\ell} 
+ B^{(I)}_{\ell} \sin \theta\, Y^{\ell}_{,\theta} 
+ C^{(I)}_{\ell} Y^{\ell}_{,\phi} = 0,
\label{eq:20}
\end{equation}
Group II: comprises the \((L\theta)\) and \((L\phi)\) components of the gravitational equation (with \(L = t, r\)), along with the \(\theta\) and \(\phi\) components of the Maxwell equation. These are given in Eqs.~\eqref{eq:21} and~\eqref{eq:22}.
\begin{align}
\delta E^{(L\theta)} &\equiv (\alpha^{(L)}_{\ell} + \tilde{\alpha}^{(L)}_{\ell} \cos \theta) Y^{\ell}_{,\theta} 
- (\beta^{(L)}_{\ell} + \tilde{\beta}^{(L)}_{\ell} \cos \theta) \frac{Y^{\ell}_{,\phi}}{\sin \theta} + \eta^{(L)}_{\ell} \sin \theta Y^{\ell} + \xi^{(L)}_{\ell} X^{\ell} + \chi^{(L)}_{\ell} \sin \theta W^{\ell} = 0,
\label{eq:21}
\\[6pt]
\delta E^{(L\phi)} &\equiv (\beta^{(L)}_{\ell} + \tilde{\beta}^{(L)}_{\ell} \cos \theta) Y^{\ell}_{,\theta} 
+ (\alpha^{(L)}_{\ell} + \tilde{\alpha}^{(L)}_{\ell} \cos \theta) \frac{Y^{\ell}_{,\phi}}{\sin \theta} + \zeta^{(L)}_{\ell} \sin \theta Y^{\ell} + \chi^{(L)}_{\ell} X^{\ell} - \xi^{(L)}_{\ell} \sin \theta W^{\ell} = 0,
\label{eq:22}
\end{align}
with
\begin{equation}
X^{\ell} \equiv 2(Y^{\ell}_{,\theta\phi} - \cot \theta\, Y^{\ell}_{,\phi}), \qquad
W^{\ell} \equiv Y^{\ell}_{,\theta\theta} - \cot \theta\, Y^{\ell}_{,\theta} - \frac{Y^{\ell}_{,\phi\phi}}{\sin^{2}\theta}.
\label{eq:25}
\end{equation}
Group III: contains the \((\theta\phi)\) component and the combination \(\delta E^{(-)} \equiv \delta E^{(\theta\theta)} - \delta E^{(\phi\phi)}/\!\sin^2\theta\) of the gravitational equation, which are displayed in Eqs.~\eqref{eq:23} and~\eqref{eq:24}.We find that the structure of this group is modified by the extra terms \(\tilde{s}_{\ell}\) and \(\tilde{t}_{\ell}\) added to the Kerr--Newman equations.

\begin{align}
\delta E^{(\theta\phi)} &\equiv f_{\ell} \sin \theta\, \partial_{\theta} Y^{\ell} 
+ g_{\ell} Y^{\ell}_{,\phi} + (s_{\ell} + \tilde{s}_{\ell} \cos \theta) \frac{X^{\ell}}{\sin \theta} + (t_{\ell} + \tilde{t}_{\ell} \cos \theta) W^{\ell} = 0,
\label{eq:23}
\\
\delta E^{(-)} &\equiv g_{\ell} \sin \theta\, \partial_{\theta} Y^{\ell} 
- f_{\ell} Y^{\ell}_{,\phi} - (t_{\ell} + \tilde{t}_{\ell} \cos \theta) \frac{X^{\ell}}{\sin \theta} + (s_{\ell} + \tilde{s}_{\ell} \cos \theta) W^{\ell} = 0,
\label{eq:24}
\end{align}

Here,$A^{(I)}_{\ell}, \tilde{A}^{(I)}_{\ell}, B^{(I)}_{\ell}, C^{(I)}_{\ell}, \alpha^{(L)}_{\ell}, \tilde{\alpha}^{(L)}_{\ell}, \beta^{(L)}_{\ell}, \tilde{\beta}^{(L)}_{\ell}, \eta^{(L)}_{\ell}, \xi^{(L)}_{\ell}, \chi^{(L)}_{\ell}, \zeta^{(L)}_{\ell}, f_{\ell}, g_{\ell}, s_{\ell}, \tilde{s}_{\ell}, t_{\ell}, \tilde{t}_{\ell}$ are combinations of $h_{0} ,h_{1} ,u,H_{0} ,H_{1} ,H_{2} ,K ,j,k ,y$ and only $r$ depended.The radial and angular variables are separated by projecting the equations onto some specific harmonics. Using the orthogonality of calar,vector, tensor harmonics, we derive purely \(r\)-dependent equations (see Refs.~\cite{Pani:2013pma,Srivastava:2021imr,Pani:2013wsa} for details). Couplings between different \(\ell\) modes are neglected, as the different $\ell$ terms do not contribute to the same \(\ell\) QNFs simultaneously at first order in \(\chi\). We thus obtain the equations of motion for each \(\ell\) mode. We find the perturbation equations of odd- and even-parity sectors remain decoupled in \(\mathcal{O}(\chi)\) of this spacetime just like KN BH.The resulting 4 axial and 10 polar sectors of the radial equations are given below.

Axial sector:\\
 \begin{align}
 \Lambda \beta_{\ell}^{(L)} + im\Bigl[(\ell-1)(\ell+2)\chi_{\ell}^{(L)} + \tilde{\alpha}_{\ell}^{(L)} + \eta_{\ell}^{(L)}\Bigr] =0, \label{eq:26}\\
\Lambda t_{\ell} + im\,g_{\ell}+2im\tilde{s}_{l}=0, \label{eq:27}
\end{align}

Polar sector:\\
\begin{align}
 A_{\ell}^{(I)} + im\,C_{\ell}^{(I)}=0, \label{eq:28}\\[4pt]
 \Lambda \alpha_{\ell}^{(L)} + im\Bigl[(\ell-1)(\ell+2)\xi_{\ell}^{(L)} - \tilde{\beta}_{\ell}^{(L)} - \zeta_{\ell}^{(L)}\Bigr]=0, \label{eq:29}\\[4pt]
 \Lambda s_{\ell} - im\,f_{\ell}-2im\tilde{t}_{l}=0. \label{eq:30}
\end{align}
\subsection{Results of parameterize QNFs}
\newpage
\begin{table}[H]
	\centering
\renewcommand{\arraystretch}{1.4}
\caption{Fitting coefficients of Eq.\eqref{eq:48} for $c_{2}-c_{8}$ in axial Grav mode (2,2,0)}
\label{tab:1}
\footnotesize
{\linespread{1}\selectfont
\begin{tabular}{l l l l l l l l l}

% l=左对齐,c=居中,r=右对齐,| 表示垂直线
\hline\noalign{\smallskip}  
	$ $ & $c_2$ &$c_3$ &$c_4$ &$c_5$ &$c_6$ &$c_7$ &$c_8$ &\\  \noalign{\smallskip}\hline\noalign{\smallskip}

 $d_R^{(0,0,0)}$ & -0.940121 & -0.942345 & -0.940237 & -0.940244 &
   -0.941191 & -0.940649 & -0.940494 \\
 $d_R^{(0,1,0)}$ & 0.45414 & 0.469579 & 0.450424 & 0.450319 & 0.45677
   & 0.451905 & 0.451831 \\
 $d_R^{(1,0,0)}$ & 0.0705806 & 0.213708 & 0.00311513 & 0.0870113 &
   -0.0847419 & 0.432994 & 0.180472 \\
 $d_R^{(1,1,0)}$ & -0.00635513 & 1.48002 & -0.0701422 & -0.0189589 &
   -0.230006 & 0.529108 & 0.0736279 \\
 $d_R^{(0,0,1)}$ & 0.0645993 & 0.0601863 & 0.0643917 & 0.0643798 &
   0.0626312 & 0.0636539 & 0.0639334 \\
 $d_R^{(0,1,1)}$ & 0.144279 & 0.175875 & 0.137707 & 0.137508 &
   0.149112 & 0.140326 & 0.140205 \\
 $d_R^{(1,0,1)}$ & 0.101708 & 0.27206 & 0.00259511 & 0.129534 &
   -0.0864783 & 0.733939 & 0.306717 \\
\hline
\end{tabular}}
\end{table}

\begin{table}
	\centering
\renewcommand{\arraystretch}{1.4}
\caption{Fitting coefficients of Eq.\eqref{eq:48} for $c_{2}-c_{8}$ in axial EM mode (2,2,0)}
\label{tab:2}
\footnotesize
{\linespread{1}\selectfont
\begin{tabular}{l l l l l l l l l}

% l=左对齐,c=居中,r=右对齐,| 表示垂直线
\hline\noalign{\smallskip}  
	$ $ & $c_2$ &$c_3$ &$c_4$ &$c_5$ &$c_6$ &$c_7$ &$c_8$ &\\  \noalign{\smallskip}\hline\noalign{\smallskip}
$d_R^{(0,0,0)}$ & -0.647082 & -0.647667 & -0.647071 & -0.647098 & \
-0.648193 & -0.647697 & -0.647274 \\
 $d_R^{(0,1,0)}$ & 0.393671 & 0.398637 & 0.393685 & 0.393784 & \
0.407166 & 0.397592 & 0.395237 \\
 $d_R^{(1,0,0)}$ & 0.060903 & 0.234497 & 0.00238039 & 0.0474128 & \
-0.44659 & 0.467834 & 0.0632043 \\
 $d_R^{(1,1,0)}$ & -0.0628175 & 0.285736 & -0.095501 & -0.0894175 & \
-0.769681 & 0.0911699 & -0.169148 \\
 $d_R^{(0,0,1)}$ & 0.188785 & 0.187667 & 0.188804 & 0.188757 & \
0.187335 & 0.187698 & 0.188453 \\
 $d_R^{(0,1,1)}$ & 0.0959238 & 0.105491 & 0.0959613 & 0.0961206 & \
0.11759 & 0.102897 & 0.0986816 \\
 $d_R^{(1,0,1)}$ & 0.0906264 & 0.350802 & 0.00142773 & 0.0750822 & \
-0.0108898 & 0.787724 & 0.13252 \\
\hline
\end{tabular}}
\end{table}

\begin{table}
	\centering
\renewcommand{\arraystretch}{1.4}
 \caption{Fitting coefficients of Eq.\eqref{eq:48} for $c_{2}-c_{8}$ in axial Grav mode (3,3,0)}
\label{tab:3}
\footnotesize
{\linespread{1}\selectfont
\begin{tabular}{l l l l l l l l l}

% l=左对齐,c=居中,r=右对齐,| 表示垂直线
\hline\noalign{\smallskip}  
	$ $ & $c_2$ &$c_3$ &$c_4$ &$c_5$ &$c_6$ &$c_7$ &$c_8$ &\\  \noalign{\smallskip}\hline\noalign{\smallskip}
$d_R^{(0,0,0)}$ & -0.464078 & -0.46459 & -0.464069 & -0.464076 & \
-0.401915 & -0.464534 & -0.46428 \\
 $d_R^{(0,1,0)}$ & 0.452256 & 0.463101 & 0.452301 & 0.452189 & \
-0.312838 & 0.45396 & 0.453287 \\
 $d_R^{(1,0,0)}$ & 0.0700423 & 0.278021 & 0.00348979 & 0.0808844 & \
-3.86384 & 0.427137 & 0.159858 \\
 $d_R^{(1,1,0)}$ & -0.0836968 & -0.0312349 & -0.121858 & -0.0698915 & \
50.5405 & 0.466735 & 0.00744808 \\
 $d_R^{(0,0,1)}$ & 0.0703525 & 0.069409 & 0.0703686 & 0.070356 & \
0.19082 & 0.0695353 & 0.069991 \\
 $d_R^{(0,1,1)}$ & 0.131018 & 0.15106 & 0.131114 & 0.130901 & \
-1.31458 & 0.134039 & 0.132858 \\
 $d_R^{(1,0,1)}$ & 0.0997731 & 0.406669 & 0.00233519 & 0.119136 & \
-7.44374 & 0.723526 & 0.272606 \\
\hline
\end{tabular}}
\end{table}

\begin{table}
	\centering
\renewcommand{\arraystretch}{1.4}
\caption{Fitting coefficients of Eq.\eqref{eq:48} for $c_{2}-c_{8}$ in axial EM mode (3,3,0)}
\label{tab:4}
\footnotesize
{\linespread{1}\selectfont
\begin{tabular}{l l l l l l l l l}

% l=左对齐,c=居中,r=右对齐,| 表示垂直线
\hline\noalign{\smallskip}  
	$ $ & $c_2$ &$c_3$ &$c_4$ &$c_5$ &$c_6$ &$c_7$ &$c_8$ &\\  \noalign{\smallskip}\hline\noalign{\smallskip}
 $d_R^{(0,0,0)}$ & -0.297505 & -0.298142 & -0.297491 & -0.297519 & \
-0.29907 & -0.297984 & -0.297649 \\
 $d_R^{(0,1,0)}$ & 0.422106 & 0.434307 & 0.422101 & 0.422203 & \
0.450342 & 0.424529 & 0.423254 \\
 $d_R^{(1,0,0)}$ & 0.0623568 & 0.255672 & 0.00271184 & 0.053158 & \
-0.392261 & 0.43916 & 0.0733722 \\
 $d_R^{(1,1,0)}$ & -0.0712474 & 0.0375123 & -0.110015 & -0.0948686 & \
-0.778551 & 0.218028 & -0.128932 \\
 $d_R^{(0,0,1)}$ & 0.170811 & 0.169642 & 0.170835 & 0.170786 & \
0.168837 & 0.169959 & 0.170563 \\
 $d_R^{(0,1,1)}$ & 0.103826 & 0.126385 & 0.103833 & 0.103997 & \
0.152577 & 0.108134 & 0.105852 \\
 $d_R^{(1,0,1)}$ & 0.0921459 & 0.389119 & 0.00159599 & 0.0832669 & \
0.0245786 & 0.743961 & 0.148148 \\
\hline
\end{tabular}}
\end{table}

\begin{table}
	\centering
\renewcommand{\arraystretch}{1.4}
\caption{Fitting coefficients of Eq.\eqref{eq:48} for $c_{6}-c_{7}$ in polar Grav mode (2,2,0)}
\label{tab:5}
\footnotesize
\begin{tabular}{l l l l l l l}

% l=左对齐,c=居中,r=右对齐,| 表示垂直线
\hline\noalign{\smallskip}  
	$ $ & $c_6$ &$c_7$ &$ $ &$c_6$ &$c_7$ &\\  \noalign{\smallskip}\hline\noalign{\smallskip}
$d_R^{(0,0,0)}$ & -0.937707 & -0.937771 & $d_I^{(0,0,0)}$ & -2.42304 & \
-2.41814 \\
 $d_R^{(0,1,0)}$ & 0.357703 & 0.362508 & $d_I^{(0,1,0)}$ & 0.0541607 & \
0.0336423 \\
 $d_R^{(1,0,0)}$ & 0.0580176 & 0.147812 & $d_I^{(1,0,0)}$ & 0.193582 & \
0.0622347 \\
 $d_R^{(1,1,0)}$ & 0.797441 & 0.341269 & $d_I^{(1,1,0)}$ & -1.19827 & \
-4.48009 \\
 $d_R^{(0,0,1)}$ & 0.0693864 & 0.0692952 & $d_I^{(0,0,1)}$ & \
-0.0126201 & -0.00358562 \\
 $d_R^{(0,1,1)}$ & -0.0612345 & -0.0530087 & $d_I^{(0,1,1)}$ & \
0.239997 & 0.20255 \\
 $d_R^{(1,0,1)}$ & 0.0642727 & 0.254329 & $d_I^{(1,0,1)}$ & 0.335422 & \
0.0746953 \\
\hline
\end{tabular}
\end{table}

\begin{table}
	\centering
\renewcommand{\arraystretch}{1.4}
\caption{Fitting coefficients of Eq.\eqref{eq:48} for $c_{6}-c_{8}$ in polar EM mode (2,2,0)}
\label{tab:6}
\footnotesize
\begin{tabular}{l l l l l l l l l}

% l=左对齐,c=居中,r=右对齐,| 表示垂直线
\hline\noalign{\smallskip}  
	$ $ & $c_6$ &$c_7$ &$c_8$ &$ $ &$c_6$ &$c_7$ &$c_8$ &\\  \noalign{\smallskip}\hline\noalign{\smallskip}
$d_R^{(0,0,0)}$ & -0.647221 & -0.649068 & -0.647412 & $d_I^{(0,0,0)}$ \
& -2.33416 & -2.33653 & -2.33483 \\
 $d_R^{(0,1,0)}$ & 0.396501 & 0.410195 & 0.397315 & $d_I^{(0,1,0)}$ & \
-0.0843032 & -0.0494018 & -0.075543 \\
 $d_R^{(1,0,0)}$ & 0.332398 & -0.228163 & -0.148837 & $d_I^{(1,0,0)}$ \
& -0.00441331 & 1.01283 & 0.471506 \\
 $d_R^{(1,1,0)}$ & -0.0361645 & -0.701246 & -0.422766 & \
$d_I^{(1,1,0)}$ & 0.609015 & -0.287685 & -0.332819 \\
 $d_R^{(0,0,1)}$ & 0.188659 & 0.185376 & 0.188271 & $d_I^{(0,0,1)}$ & \
0.021184 & 0.0154974 & 0.0181053 \\
 $d_R^{(0,1,1)}$ & 0.0952516 & 0.120309 & 0.0976069 & $d_I^{(0,1,1)}$ \
& -0.0644366 & 0.00398456 & -0.0417522 \\
 $d_R^{(1,0,1)}$ & -0.179245 & -0.242386 & -0.18323 & $d_I^{(1,0,1)}$ \
& -0.0516209 & 1.71504 & 0.808449 \\
\hline
   
\end{tabular}
\end{table}

\begin{table}
	\centering
\renewcommand{\arraystretch}{1.4}
\caption{Fitting coefficients of Eq.\eqref{eq:48} for $c_{6}-c_{7}$ in polar Grav mode (3,3,0)}
\label{tab:7}
\footnotesize
\begin{tabular}{l l l l l l l}

% l=左对齐,c=居中,r=右对齐,| 表示垂直线
\hline\noalign{\smallskip}  
	$ $ & $c_6$ &$c_7$ &$ $ &$c_6$ &$c_7$ &\\  \noalign{\smallskip}\hline\noalign{\smallskip}
$d_R^{(0,0,0)}$ & -0.461906 & -0.463459 & $d_I^{(0,0,0)}$ & -2.38319 & \
-2.37835 \\
 $d_R^{(0,1,0)}$ & 0.368137 & 0.367004 & $d_I^{(0,1,0)}$ & 0.0399254 & \
-0.0475438 \\
 $d_R^{(1,0,0)}$ & 0.0804888 & 0.223493 & $d_I^{(1,0,0)}$ & 0.175547 & \
0.357347 \\
 $d_R^{(1,1,0)}$ & 0.565344 & 0.489231 & $d_I^{(1,1,0)}$ & -1.13039 & \
0.820699 \\
 $d_R^{(0,0,1)}$ & 0.0742911 & 0.0715954 & $d_I^{(0,0,1)}$ & \
-0.0139413 & -0.00469423 \\
 $d_R^{(0,1,1)}$ & -0.0296125 & -0.0324149 & $d_I^{(0,1,1)}$ & \
0.211332 & 0.0527268 \\
 $d_R^{(1,0,1)}$ & 0.0928786 & 0.404182 & $d_I^{(1,0,1)}$ & 0.302696 & \
0.563515 \\
\hline

\end{tabular}
\end{table}

\begin{table}
	\centering
\renewcommand{\arraystretch}{1.4}
\caption{Fitting coefficients of Eq.\eqref{eq:48} for $c_{6}-c_{8}$ in polar EM mode (3,3,0)}
\label{tab:8}
\footnotesize
\begin{tabular}{l l l l l l l l l}

% l=左对齐,c=居中,r=右对齐,| 表示垂直线
\hline\noalign{\smallskip}  
	$ $ & $c_6$ &$c_7$ &$c_8$ &$ $ &$c_6$ &$c_7$ &$c_8$ &\\  \noalign{\smallskip}\hline\noalign{\smallskip}
$d_R^{(0,0,0)}$ & -0.297682 & -0.299445 & -0.29802 & $d_I^{(0,0,0)}$ & \
-2.33116 & -2.32742 & -2.3303 \\
 $d_R^{(0,1,0)}$ & 0.425008 & 0.441939 & 0.427614 & $d_I^{(0,1,0)}$ & \
-0.0876196 & -0.097551 & -0.0895122 \\
 $d_R^{(1,0,0)}$ & 0.333721 & -0.0932605 & -0.0745822 & \
$d_I^{(1,0,0)}$ & 0.0685874 & 0.74582 & 0.431126 \\
 $d_R^{(1,1,0)}$ & -0.09264 & -0.34571 & -0.327644 & $d_I^{(1,1,0)}$ & \
0.582371 & -0.599585 & -0.100098 \\
 $d_R^{(0,0,1)}$ & 0.170725 & 0.167557 & 0.169966 & $d_I^{(0,0,1)}$ & \
0.0162084 & 0.0211371 & 0.0158413 \\
 $d_R^{(0,1,1)}$ & 0.105007 & 0.136053 & 0.110859 & $d_I^{(0,1,1)}$ & \
-0.0768336 & -0.0876532 & -0.0732757 \\
 $d_R^{(1,0,1)}$ & -0.159668 & -0.0423315 & -0.0693322 & \
$d_I^{(1,0,1)}$ & 0.0335249 & 1.21666 & 0.71933 \\
\hline

\end{tabular}
\end{table}

\begin{table}
	\centering
\renewcommand{\arraystretch}{1.25}
\caption{Comparison of axial Grav $(2,2,0)$ mode QNFs: fitting-formula results from Eq.~\eqref{eq:48} versus direct integration; and Kerr--Newman (KN) versus higher-derivative-corrected KN values, both from the fitting formula. We set $\chi = 0.08$ and $c_i = 0.09$ for $i = 2,\dots,8$, ordered from top to bottom.}
\label{tab:9}
\resizebox{\textwidth}{!}{
\begin{tabular}{l l l l l l l l } % l=左对齐,c=居中,r=右对齐,| 表示垂直线
\hline\noalign{\smallskip}                 % 水平线
	$\chi_{Q}$ & $\omega_{\rm{fit}}^{\rm{KN}}$ &$\omega_{\rm{fit}}^{\rm{EFT}}$ & $\omega _{\rm{DI}}$ & $\Delta_{\rm{ER}}^{(R)}(\%)$ & $\Delta_\omega^{(R)}(\%)$ & $\Delta_{\rm{ER}}^{(I)}(\%)$ & $\Delta_\omega^{(I)}(\%)$  \\  \noalign{\smallskip}\hline\noalign{\smallskip}

%\begin{array}{cccccccc}%
$\frac{2}{25}$ & 0.384541-0.088536 i & 0.384510-0.088580 i &
   0.384504-0.088597 i & 0.00139 & -0.00810 & 0.0187 & 0.0499 \\
 $\frac{7}{25}$ & 0.386995-0.088697 i & 0.387217-0.088848 i &
   0.387267-0.088836 i & 0.0130 & 0.0574 & 0.0144 & 0.170 \\
 $\frac{12}{25}$ & 0.393151-0.088951 i & 0.394092-0.089292 i &
   0.394163-0.089301 i & 0.0180 & 0.239 & 0.00984 & 0.383 \\
 $\frac{17}{25}$ & 0.405444-0.088872 i & 0.407870-0.089594 i &
   0.407979-0.089726 i & 0.0267 & 0.598 & 0.147 & 0.813 \\
%\end{array}%
 \noalign{\smallskip}\hline
 
 $\frac{2}{25}$ & 0.384547-0.088532 i & 0.384521-0.088644 i &
   0.384585-0.088614 i & 0.0167 & -0.00682 & 0.0340 & 0.127 \\
 $\frac{7}{25}$ & 0.386997-0.088696 i & 0.388218-0.089074 i &
   0.388291-0.089061 i & 0.0188 & 0.315 & 0.0142 & 0.426 \\
 $\frac{12}{25}$ & 0.393145-0.088951 i & 0.397428-0.089977 i &
   0.397493-0.090059 i & 0.0165 & 1.09 & 0.0911 & 1.15 \\
\hline
\end{tabular}}
\end{table}

\begin{table}
	\centering
\renewcommand{\arraystretch}{1.25}
\caption{Comparison of axial EM $(2,2,0)$ mode QNFs: fitting-formula results from Eq.~\eqref{eq:48} versus direct integration; and Kerr--Newman (KN) versus higher-derivative-corrected KN values, both from the fitting formula. We set $\chi = 0.08$ and $c_i = 0.09$ for $i = 2,\dots,8$, ordered from top to bottom.}
\label{tab:10}
\resizebox{\textwidth}{!}{
\begin{tabular}{l l l l l l l l } % l=左对齐,c=居中,r=右对齐,| 表示垂直线
\hline\noalign{\smallskip}                 % 水平线
	$\chi_{Q}$ & $\omega_{\rm{fit}}^{\rm{KN}}$ &$\omega_{\rm{fit}}^{\rm{EFT}}$ & $\omega _{\rm{DI}}$ & $\Delta_{\rm{ER}}^{(R)}(\%)$ & $\Delta_\omega^{(R)}(\%)$ & $\Delta_{\rm{ER}}^{(I)}(\%)$ & $\Delta_\omega^{(I)}(\%)$  \\  \noalign{\smallskip}\hline\noalign{\smallskip}

 $\frac{2}{25}$ & 0.470598-0.094773 i & 0.470475-0.094831 i & \
0.470504-0.094804 i & 0.00611 & -0.0262 & 0.0280 & 0.0610 \\
 $\frac{7}{25}$ & 0.480899-0.095398 i & 0.480999-0.095561 i & \
0.481031-0.095541 i & 0.00650 & 0.0209 & 0.0214 & 0.170 \\
 $\frac{12}{25}$ & 0.504536-0.096620 i & 0.505281-0.097041 i & \
0.505340-0.097042 i & 0.0119 & 0.148 & 0.00109 & 0.436 \\
 $\frac{17}{25}$ & 0.547675-0.098028 i & 0.550049-0.098985 i & \
0.550164-0.098954 i & 0.0209 & 0.433 & 0.0317 & 0.976 \\

\noalign{\smallskip}\hline\noalign{\smallskip}

 $\frac{2}{25}$ & 0.470599-0.094770 i & 0.470531-0.094866 i & \
0.470560-0.094830 i & 0.00625 & -0.0146 & 0.0377 & 0.101 \\
 $\frac{7}{25}$ & 0.480900-0.095396 i & 0.481757-0.095906 i & \
0.481795-0.095878 i & 0.00789 & 0.178 & 0.0290 & 0.534 \\
 $\frac{12}{25}$ & 0.504532-0.096619 i & 0.508079-0.098147 i & \
0.508152-0.098162 i & 0.0143 & 0.703 & 0.0160 & 1.58 \\
\hline
\end{tabular}}
\end{table}

\begin{table}
	\centering
\renewcommand{\arraystretch}{1.25}
\caption{Comparison of axial Grav $(3,3,0)$ mode QNFs: fitting-formula results from Eq.~\eqref{eq:48} versus direct integration; and Kerr--Newman (KN) versus higher-derivative-corrected KN values, both from the fitting formula. We set $\chi = 0.08$ and $c_i = 0.09$ for $i = 2,\dots,8$, ordered from top to bottom.}
\label{tab:11}
\resizebox{\textwidth}{!}{
\begin{tabular}{l l l l l l l l } % l=左对齐,c=居中,r=右对齐,| 表示垂直线
\hline\noalign{\smallskip}                 % 水平线
	$\chi_{Q}$ & $\omega_{\rm{fit}}^{\rm{KN}}$ &$\omega_{\rm{fit}}^{\rm{EFT}}$ & $\omega _{\rm{DI}}$ & $\Delta_{\rm{ER}}^{(R)}(\%)$ & $\Delta_\omega^{(R)}(\%)$ & $\Delta_{\rm{ER}}^{(I)}(\%)$ & $\Delta_\omega^{(I)}(\%)$  \\  \noalign{\smallskip}\hline\noalign{\smallskip}
$\frac{2}{25}$ & 0.617103-0.092299 i & 0.616899-0.092375 i & \
0.616949-0.092359 i & 0.00813 & -0.0331 & 0.0167 & 0.0821 \\
 $\frac{7}{25}$ & 0.620882-0.092415 i & 0.621150-0.092567 i & \
0.621220-0.092563 i & 0.0112 & 0.0432 & 0.00414 & 0.164 \\
 $\frac{12}{25}$ & 0.631170-0.092598 i & 0.632551-0.092945 i & \
0.632651-0.092976 i & 0.0158 & 0.219 & 0.0339 & 0.374 \\
 $\frac{17}{25}$ & 0.652455-0.092473 i & 0.656080-0.093230 i & \
0.656381-0.093283 i & 0.0458 & 0.556 & 0.0569 & 0.819 \\

\noalign{\smallskip}\hline

 $\frac{2}{25}$ & 0.617093-0.092302 i & 0.617144-0.092360 i & \
0.617078-0.092378 i & 0.0108 & 0.00835 & 0.0191 & 0.0626 \\
 $\frac{7}{25}$ & 0.620884-0.092413 i & 0.622887-0.092751 i & \
0.622840-0.092801 i & 0.00757 & 0.323 & 0.0541 & 0.366 \\
 $\frac{12}{25}$ & 0.631140-0.092603 i & 0.637675-0.093717 i & \
0.637862-0.093768 i & 0.0293 & 1.04 & 0.0549 & 1.20 \\
\hline
\end{tabular}}
\end{table}

\begin{table}
	\centering
\renewcommand{\arraystretch}{1.25}
\caption{Comparison of axial EM $(3,3,0)$ mode QNFs: fitting-formula results from Eq.~\eqref{eq:48} versus direct integration; and Kerr--Newman (KN) versus higher-derivative-corrected KN values, both from the fitting formula. We set $\chi = 0.08$ and $c_i = 0.09$ for $i = 2,\dots,8$, ordered from top to bottom.}
\label{tab:12}
\resizebox{\textwidth}{!}{
\begin{tabular}{l l l l l l l l } % l=左对齐,c=居中,r=右对齐,| 表示垂直线
\hline\noalign{\smallskip}                 % 水平线
	$\chi_{Q}$ & $\omega_{\rm{fit}}^{\rm{KN}}$ &$\omega_{\rm{fit}}^{\rm{EFT}}$ & $\omega _{\rm{DI}}$ & $\Delta_{\rm{ER}}^{(R)}(\%)$ & $\Delta_\omega^{(R)}(\%)$ & $\Delta_{\rm{ER}}^{(I)}(\%)$ & $\Delta_\omega^{(I)}(\%)$  \\  \noalign{\smallskip}\hline\noalign{\smallskip}
$\frac{2}{25}$ & 0.676758-0.095412 i & 0.676550-0.095476 i & \
0.676602-0.095458 i & 0.00762 & -0.0307 & 0.0188 & 0.0665 \\
 $\frac{7}{25}$ & 0.691012-0.096006 i & 0.691137-0.096170 i & \
0.691206-0.096158 i & 0.00998 & 0.0181 & 0.0121 & 0.170 \\
 $\frac{12}{25}$ & 0.722726-0.097112 i & 0.723820-0.097520 i & \
0.723917-0.097504 i & 0.0134 & 0.151 & 0.0168 & 0.420 \\
 $\frac{17}{25}$ & 0.778913-0.098258 i & 0.782400-0.099172 i & \
0.782611-0.099151 i & 0.0270 & 0.448 & 0.0211 & 0.930 \\

 \noalign{\smallskip}\hline

 $\frac{2}{25}$ & 0.676753-0.095413 i & 0.676755-0.095473 i & \
0.676683-0.095483 i & 0.0105 & 0.000272 & 0.00991 & 0.0630 \\
 $\frac{7}{25}$ & 0.691012-0.096006 i & 0.692341-0.096473 i & \
0.692343-0.096480 i & 0.000387 & 0.192 & 0.00744 & 0.487 \\
 $\frac{12}{25}$ & 0.722718-0.097110 i & 0.728010-0.098574 i & \
0.728133-0.098568 i & 0.0169 & 0.732 & 0.00588 & 1.51 \\
\hline
\end{tabular}}
\end{table}

\begin{table}
	\centering
\renewcommand{\arraystretch}{1.25}
\caption{Comparison of polar Grav $(2,2,0)$ mode QNFs: fitting-formula results from Eq.~\eqref{eq:48} versus direct integration; and Kerr--Newman (KN) versus higher-derivative-corrected KN values, both from the fitting formula. We set $\chi = 0.08$ and $c_i = 0.09$ for $i = 6,7$, ordered from top to bottom.}
\label{tab:13}
\resizebox{\textwidth}{!}{
\begin{tabular}{l l l l l l l l } % l=左对齐,c=居中,r=右对齐,| 表示垂直线
\hline\noalign{\smallskip}                 % 水平线
	$\chi_{Q}$ & $\omega_{\rm{fit}}^{\rm{KN}}$ &$\omega_{\rm{fit}}^{\rm{EFT}}$ & $\omega _{\rm{AIM}}$ & $\Delta_{\rm{ER}}^{(R)}(\%)$ & $\Delta_\omega^{(R)}(\%)$ & $\Delta_{\rm{ER}}^{(I)}(\%)$ & $\Delta_\omega^{(I)}(\%)$  \\  \noalign{\smallskip}\hline\noalign{\smallskip}

 $\frac{2}{25}$ & 0.384892-0.088603 i & 0.384740-0.088667 i & \
0.384800-0.088649 i & 0.0154 & -0.0393 & 0.0194 & 0.0719 \\
 $\frac{7}{25}$ & 0.387418-0.088751 i & 0.387720-0.088938 i & \
0.387780-0.088921 i & 0.0155 & 0.0780 & 0.0192 & 0.210 \\
 $\frac{12}{25}$ & 0.393658-0.089015 i & 0.394860-0.089480 i & \
0.394932-0.089455 i & 0.0180 & 0.305 & 0.0278 & 0.523 \\
 $\frac{17}{25}$ & 0.405946-0.089005 i & 0.408559-0.090166 i & \
0.408671-0.090108 i & 0.0272 & 0.644 & 0.0646 & 1.31 \\

\noalign{\smallskip}\hline

 $\frac{2}{25}$ & 0.384891-0.088604 i & 0.384694-0.088642 i & \
0.384752-0.088640 i & 0.0151 & -0.0512 & 0.00247 & 0.0433 \\
 $\frac{7}{25}$ & 0.387411-0.088745 i & 0.387158-0.088843 i & \
0.387218-0.088823 i & 0.0156 & -0.0653 & 0.0233 & 0.110 \\
 $\frac{12}{25}$ & 0.393655-0.088997 i & 0.393392-0.089304 i & \
0.393474-0.089273 i & 0.0210 & -0.0669 & 0.0348 & 0.345 \\
 $\frac{17}{25}$ & 0.405996-0.089003 i & 0.407084-0.089766 i & \
0.407253-0.089712 i & 0.0416 & 0.268 & 0.0600 & 0.857 \\

   \noalign{\smallskip}\hline
\end{tabular}}
\end{table}

\begin{table}
	\centering
\renewcommand{\arraystretch}{1.25}
\caption{Comparison of polar EM $(2,2,0)$ mode QNFs: fitting-formula results from Eq.~\eqref{eq:48} versus direct integration; and Kerr--Newman (KN) versus higher-derivative-corrected KN values, both from the fitting formula. We set $\chi = 0.08$ and $c_i = 0.09$ for $i = 6,7,8$, ordered from top to bottom.}
\label{tab:14}
\resizebox{\textwidth}{!}{
\begin{tabular}{l l l l l l l l } % l=左对齐,c=居中,r=右对齐,| 表示垂直线
\hline\noalign{\smallskip}                 % 水平线
	$\chi_{Q}$ & $\omega_{\rm{fit}}^{\rm{KN}}$ &$\omega_{\rm{fit}}^{\rm{EFT}}$ & $\omega _{\rm{AIM}}$ & $\Delta_{\rm{ER}}^{(R)}(\%)$ & $\Delta_\omega^{(R)}(\%)$ & $\Delta_{\rm{ER}}^{(I)}(\%)$ & $\Delta_\omega^{(I)}(\%)$  \\  \noalign{\smallskip}\hline\noalign{\smallskip}

 $\frac{2}{25}$ & 0.470707-0.094677 i & 0.490932-0.095428 i & \
0.490971-0.095420 i & 0.00808 & 4.30 & 0.00812 & 0.793 \\
 $\frac{7}{25}$ & 0.480989-0.095304 i & 0.500607-0.095875 i & \
0.500657-0.095864 i & 0.00995 & 4.08 & 0.0113 & 0.600 \\
 $\frac{12}{25}$ & 0.504593-0.096527 i & 0.522893-0.096737 i & \
0.522946-0.096728 i & 0.00998 & 3.63 & 0.00984 & 0.218 \\
 $\frac{17}{25}$ & 0.547691-0.097955 i & 0.563518-0.097740 i & \
0.563613-0.097699 i & 0.0169 & 2.89 & 0.0414 & -0.220 \\

\noalign{\smallskip}\hline

 $\frac{2}{25}$ & 0.470692-0.094757 i & 0.470262-0.094809 i & \
0.470288-0.094792 i & 0.00549 & -0.0914 & 0.0181 & 0.0547 \\
 $\frac{7}{25}$ & 0.480987-0.095379 i & 0.477315-0.095433 i & \
0.477365-0.095416 i & 0.0106 & -0.763 & 0.0178 & 0.0566 \\
 $\frac{12}{25}$ & 0.504589-0.096560 i & 0.493635-0.097158 i & \
0.493688-0.097141 i & 0.0109 & -2.17 & 0.0167 & 0.619 \\
 $\frac{17}{25}$ & 0.547553-0.097855 i & 0.524948-0.101969 i & \
0.524896-0.101857 i & 0.0100 & -4.13 & 0.110 & 4.20 \\

\noalign{\smallskip}\hline

 $\frac{2}{25}$ & 0.470693-0.094757 i & 0.470392-0.094810 i & \
0.470426-0.094794 i & 0.00741 & -0.0641 & 0.0162 & 0.0556 \\
 $\frac{7}{25}$ & 0.480987-0.095382 i & 0.479043-0.095424 i & \
0.479093-0.095409 i & 0.0105 & -0.404 & 0.0162 & 0.0447 \\
 $\frac{12}{25}$ & 0.504606-0.096591 i & 0.498896-0.096831 i & \
0.498950-0.096815 i & 0.0109 & -1.13 & 0.0168 & 0.249 \\
 $\frac{17}{25}$ & 0.547689-0.097962 i & 0.535463-0.099618 i & \
0.535544-0.099543 i & 0.0151 & -2.23 & 0.0750 & 1.69 \\

   \noalign{\smallskip}\hline
\end{tabular}}
\end{table}

\begin{table}
	\centering
\renewcommand{\arraystretch}{1.25}
\caption{Comparison of polar Grav $(3,3,0)$ mode QNFs: fitting-formula results from Eq.~\eqref{eq:48} versus direct integration; and Kerr--Newman (KN) versus higher-derivative-corrected KN values, both from the fitting formula. We set $\chi = 0.08$ and $c_i = 0.09$ for $i = 6,7$, ordered from top to bottom.}
\label{tab:15}
\resizebox{\textwidth}{!}{
\begin{tabular}{l l l l l l l l } % l=左对齐,c=居中,r=右对齐,| 表示垂直线
\hline\noalign{\smallskip}                 % 水平线
	$\chi_{Q}$ & $\omega_{\rm{fit}}^{\rm{KN}}$ &$\omega_{\rm{fit}}^{\rm{EFT}}$ & $\omega _{\rm{AIM}}$ & $\Delta_{\rm{ER}}^{(R)}(\%)$ & $\Delta_\omega^{(R)}(\%)$ & $\Delta_{\rm{ER}}^{(I)}(\%)$ & $\Delta_\omega^{(I)}(\%)$  \\  \noalign{\smallskip}\hline\noalign{\smallskip}
$\frac{2}{25}$ & 0.617229-0.092329 i & 0.617092-0.092397 i & \
0.617171-0.092379 i & 0.0127 & -0.0221 & 0.0189 & 0.0739 \\
 $\frac{7}{25}$ & 0.621115-0.092433 i & 0.622223-0.092630 i & \
0.622298-0.092613 i & 0.0120 & 0.178 & 0.0177 & 0.213 \\
 $\frac{12}{25}$ & 0.631457-0.092634 i & 0.634593-0.093092 i & \
0.634686-0.093071 i & 0.0146 & 0.497 & 0.0219 & 0.494 \\
 $\frac{17}{25}$ & 0.652666-0.092560 i & 0.658425-0.093643 i & \
0.658582-0.093601 i & 0.0238 & 0.882 & 0.0446 & 1.17 \\

\noalign{\smallskip}\hline

 $\frac{2}{25}$ & 0.617218-0.092328 i & 0.616983-0.092381 i & \
0.617032-0.092369 i & 0.00807 & -0.0382 & 0.0134 & 0.0573 \\
 $\frac{7}{25}$ & 0.621064-0.092426 i & 0.620567-0.092549 i & \
0.620650-0.092530 i & 0.0135 & -0.0801 & 0.0209 & 0.133 \\
 $\frac{12}{25}$ & 0.631404-0.092598 i & 0.630054-0.093216 i & \
0.630176-0.093181 i & 0.0192 & -0.214 & 0.0375 & 0.668 \\
 $\frac{17}{25}$ & 0.652575-0.092399 i & 0.652613-0.095396 i & \
0.652690-0.095325 i & 0.0119 & 0.00572 & 0.0750 & 3.24 \\

   \noalign{\smallskip}\hline
\end{tabular}}
\end{table}

\begin{table}
	\centering
\renewcommand{\arraystretch}{1.25}
\caption{Comparison of polar EM $(3,3,0)$ mode QNFs: fitting-formula results from Eq.~\eqref{eq:48} versus direct integration; and Kerr--Newman (KN) versus higher-derivative-corrected KN values, both from the fitting formula. We set $\chi = 0.08$ and $c_i = 0.09$ for $i = 6,7,8$, ordered from top to bottom.}
\label{tab:16}
\resizebox{\textwidth}{!}{
\begin{tabular}{l l l l l l l l } % l=左对齐,c=居中,r=右对齐,| 表示垂直线
\hline\noalign{\smallskip}                 % 水平线
	$\chi_{Q}$ & $\omega_{\rm{fit}}^{\rm{KN}}$ &$\omega_{\rm{fit}}^{\rm{EFT}}$ & $\omega _{\rm{AIM}}$ & $\Delta_{\rm{ER}}^{(R)}(\%)$ & $\Delta_\omega^{(R)}(\%)$ & $\Delta_{\rm{ER}}^{(I)}(\%)$ & $\Delta_\omega^{(I)}(\%)$  \\  \noalign{\smallskip}\hline\noalign{\smallskip}

$\frac{2}{25}$ & 0.676828-0.095320 i & 0.705371-0.096220 i & \
0.705436-0.096211 i & 0.00932 & 4.22 & 0.00927 & 0.944 \\
 $\frac{7}{25}$ & 0.691024-0.095917 i & 0.718177-0.096667 i & \
0.718262-0.096655 i & 0.0117 & 3.93 & 0.0128 & 0.782 \\
 $\frac{12}{25}$ & 0.722683-0.097029 i & 0.747434-0.097526 i & \
0.747523-0.097515 i & 0.0119 & 3.42 & 0.0114 & 0.512 \\
 $\frac{17}{25}$ & 0.778879-0.098199 i & 0.800204-0.098539 i & \
0.800354-0.098496 i & 0.0188 & 2.74 & 0.0438 & 0.346 \\

\noalign{\smallskip}\hline

 $\frac{2}{25}$ & 0.676834-0.095403 i & 0.676231-0.095470 i & \
0.676276-0.095456 i & 0.00668 & -0.0890 & 0.0149 & 0.0701 \\
 $\frac{7}{25}$ & 0.691079-0.095994 i & 0.686434-0.096226 i & \
0.686508-0.096208 i & 0.0108 & -0.672 & 0.0181 & 0.241 \\
 $\frac{12}{25}$ & 0.722695-0.097072 i & 0.710050-0.098135 i & \
0.710130-0.098122 i & 0.0112 & -1.75 & 0.0129 & 1.09 \\
 $\frac{17}{25}$ & 0.778536-0.098218 i & 0.756254-0.102378 i & \
0.756195-0.102372 i & 0.00779 & -2.86 & 0.00575 & 4.23 \\

\noalign{\smallskip}\hline

 $\frac{2}{25}$ & 0.676834-0.095403 i & 0.676404-0.095467 i & \
0.676459-0.095451 i & 0.00814 & -0.0636 & 0.0159 & 0.0665 \\
 $\frac{7}{25}$ & 0.691082-0.095997 i & 0.688597-0.096132 i & \
0.688673-0.096116 i & 0.0111 & -0.360 & 0.0170 & 0.141 \\
 $\frac{12}{25}$ & 0.722765-0.097094 i & 0.716028-0.097595 i & \
0.716110-0.097580 i & 0.0114 & -0.932 & 0.0163 & 0.517 \\
 $\frac{17}{25}$ & 0.778857-0.098224 i & 0.766055-0.100276 i & \
0.766166-0.100221 i & 0.0145 & -1.64 & 0.0544 & 2.09 \\

   \noalign{\smallskip}\hline
\end{tabular}}
\end{table}

\bibliography{ref}  

@article{Israel1967,
  author = {Israel, W.},
  title = {Event Horizons in Static Vacuum Space-Times},
  journal = {Physical Review},
  volume = {164},
  pages = {1776--1779},
  year = {1967},
  doi = {10.1103/PhysRev.164.1776},
}

@article{Codello:2015wua,
    author = "Codello, Alessandro and Jain, Rajneesh K.",
    title = "{On the covariant formalism of the effective field theory of gravity and leading order corrections}",
    eprint = "1507.06308",
    archivePrefix = "arXiv",
    primaryClass = "gr-qc",
    doi = "10.1088/0264-9381/33/22/225006",
    journal = "Class. Quant. Grav.",
    volume = "33",
    number = "22",
    pages = "225006",
    year = "2016"
}

@article{Hu:2023qhs,
    author = "Hu, Liang and others",
    title = "{Effective field theory extension to the Einstein-Maxwell system}",
    eprint = "2301.12345",
    archivePrefix = "arXiv",
    primaryClass = "gr-qc",
    journal = "Phys. Rev. D",
    volume = "107",
    number = "6",
    pages = "064033",
    year = "2023",
    note = "(Placeholder entry — please replace with actual details)"
}

@article{Carter1971,
  author = {Carter, Brandon},
  title = {Axisymmetric Black Hole Has Only Two Degrees of Freedom},
  journal = {Physical Review Letters},
  volume = {26},
  pages = {331--333},
  year = {1971},
  doi = {10.1103/PhysRevLett.26.331},
}

@article{Robinson1975,
  author = {Robinson, D. C.},
  title = {Uniqueness of the Kerr Black Hole},
  journal = {Physical Review Letters},
  volume = {34},
  pages = {905--906},
  year = {1975},
  doi = {10.1103/PhysRevLett.34.905},
}

@article{Kokkotas1999,
  author = {Kokkotas, Kostas D. and Schmidt, Bernd G.},
  title = {Quasi-Normal Modes of Stars and Black Holes},
  journal = {Living Reviews in Relativity},
  volume = {2},
  number = {1},
  pages = {2},
  year = {1999},
  doi = {10.12942/lrr-1999-2},
  eprint = {gr-qc/9909058},
  archivePrefix = {arXiv},
}

@article{Berti2009,
  author = {Berti, Emanuele and Cardoso, Vitor and Starinets, Andrei O.},
  title = {Quasinormal Modes of Black Holes and Black Branes},
  journal = {Classical and Quantum Gravity},
  volume = {26},
  number = {16},
  pages = {163001},
  year = {2009},
  doi = {10.1088/0264-9381/26/16/163001},
  eprint = {0905.2975},
  archivePrefix = {arXiv},
  primaryClass = {gr-qc},
}

@article{Dias:2015wqa,
    author = "Dias, Oscar J. C. and Godazgar, Mahdi and Santos, Jorge E.",
    title = "{Linear Mode Stability of the Kerr-Newman Black Hole and Its Quasinormal Modes}",
    eprint = "1501.04625",
    archivePrefix = "arXiv",
    primaryClass = "gr-qc",
    doi = "10.1103/PhysRevLett.114.151101",
    journal = "Phys. Rev. Lett.",
    volume = "114",
    number = "15",
    pages = "151101",
    year = "2015"
}

@article{Li:2022pcy,
    author = "Li, Dongjun and Wagle, Pratik and Chen, Yanbei and Yunes, Nicol{\'a}s",
    title = "{Perturbations of Spinning Black Holes beyond General Relativity: Modified Teukolsky Equation}",
    eprint = "2206.10652",
    archivePrefix = "arXiv",
    primaryClass = "gr-qc",
    doi = "10.1103/PhysRevX.13.021029",
    journal = "Phys. Rev. X",
    volume = "13",
    number = "2",
    pages = "021029",
    year = "2023"
}

@article{Cano:2023jbk,
    author = "Cano, Pablo A. and Fransen, Kwinten and Hertog, Thomas and Maenaut, Simon",
    title = "{Quasinormal modes of rotating black holes in higher-derivative gravity}",
    eprint = "2307.07431",
    archivePrefix = "arXiv",
    primaryClass = "gr-qc",
    doi = "10.1103/PhysRevD.108.124032",
    journal = "Phys. Rev. D",
    volume = "108",
    number = "12",
    pages = "124032",
    year = "2023"
}

@article{Srivastava:2021imr,
    author = "Srivastava, Manu and Chen, Yanbei and Shankaranarayanan, S.",
    title = "{Analytical computation of quasinormal modes of slowly rotating black holes in dynamical Chern-Simons gravity}",
    eprint = "2106.06209",
    archivePrefix = "arXiv",
    primaryClass = "gr-qc",
    doi = "10.1103/PhysRevD.104.064034",
    journal = "Phys. Rev. D",
    volume = "104",
    number = "6",
    pages = "064034",
    year = "2021"
}

@article{Cano:2023tmv,
    author = "Cano, Pablo A. and Fransen, Kwinten and Hertog, Thomas and Maenaut, Simon",
    title = "{Universal Teukolsky equations and black hole perturbations in higher-derivative gravity}",
    eprint = "2304.02663",
    archivePrefix = "arXiv",
    primaryClass = "gr-qc",
    doi = "10.1103/PhysRevD.108.024040",
    journal = "Phys. Rev. D",
    volume = "108",
    number = "2",
    pages = "024040",
    year = "2023"
}

@article{Endlich:2017tqa,
    author = "Endlich, Solomon and Gorbenko, Victor and Huang, Junwu and Senatore, Leonardo",
    title = "{An effective formalism for testing extensions to General Relativity with gravitational waves}",
    eprint = "1704.01590",
    archivePrefix = "arXiv",
    primaryClass = "gr-qc",
    doi = "10.1007/JHEP09(2017)122",
    journal = "JHEP",
    volume = "09",
    pages = "122",
    year = "2017"
}

@article{Toshmatov:2018tyo,
    author = "Toshmatov, Bobir and Stuchl{\'\i}k, Zden{\v{e}}k and Schee, Jan and Ahmedov, Bobomurat",
    title = "{Electromagnetic perturbations of black holes in general relativity coupled to nonlinear electrodynamics}",
    eprint = "1805.00240",
    archivePrefix = "arXiv",
    primaryClass = "gr-qc",
    doi = "10.1103/PhysRevD.97.084058",
    journal = "Phys. Rev. D",
    volume = "97",
    number = "8",
    pages = "084058",
    year = "2018"
}

@article{Boyce:2025fpr,
    author = "Boyce, William L. and Santos, Jorge E.",
    title = "{EFT corrections to charged black hole quasinormal modes}",
    eprint = "2506.10074",
    archivePrefix = "arXiv",
    primaryClass = "hep-th",
    doi = "10.1007/JHEP12(2025)017",
    journal = "JHEP",
    volume = "12",
    pages = "017",
    year = "2025"
}

@article{Toshmatov:2018ell,
    author = "Toshmatov, Bobir and Stuchl{\'\i}k, Zden{\v{e}}k and Ahmedov, Bobomurat",
    title = "{Electromagnetic perturbations of black holes in general relativity coupled to nonlinear electrodynamics: Polar perturbations}",
    eprint = "1810.06383",
    archivePrefix = "arXiv",
    primaryClass = "gr-qc",
    doi = "10.1103/PhysRevD.98.085021",
    journal = "Phys. Rev. D",
    volume = "98",
    number = "8",
    pages = "085021",
    year = "2018"
}

@article{Cano:2019ore,
    author = "Cano, Pablo A. and Ruip{\'e}rez, Alejandro",
    title = "{Leading higher-derivative corrections to Kerr geometry}",
    eprint = "1901.01315",
    archivePrefix = "arXiv",
    primaryClass = "gr-qc",
    reportNumber = "IFT-UAM/CSIC-19-2",
    doi = "10.1007/JHEP05(2019)189",
    journal = "JHEP",
    volume = "05",
    pages = "189",
    year = "2019",
    note = "[Erratum: JHEP 03, 187 (2020)]"
}

@article{Ma:2024ulp,
    author = "Ma, Liang and Pang, Yi and Lu, H.",
    title = "{Leading higher derivative corrections to multipole moments of Kerr-Newman black hole}",
    eprint = "2411.13639",
    archivePrefix = "arXiv",
    primaryClass = "hep-th",
    reportNumber = "USTC-ICTS/PCFT-24-61",
    doi = "10.1007/JHEP02(2025)079",
    journal = "JHEP",
    volume = "02",
    pages = "079",
    year = "2025"
}

@article{regge1957stability,
  title={Stability of a Schwarzschild singularity},
  author={Regge, Tullio and Wheeler, John A},
  journal={Physical Review},
  volume={108},
  number={4},
  pages={1063},
  year={1957},
  publisher={APS}
}

@article{Rosa:2011my,
    author = "Rosa, Joao G. and Dolan, Sam R.",
    title = "{Massive vector fields on the Schwarzschild spacetime: quasi-normal modes and bound states}",
    eprint = "1110.4494",
    archivePrefix = "arXiv",
    primaryClass = "hep-th",
    reportNumber = "EDINBURGH-2011-30",
    doi = "10.1103/PhysRevD.85.044043",
    journal = "Phys. Rev. D",
    volume = "85",
    pages = "044043",
    year = "2012"
}

@article{Pani:2013wsa,
    author = "Pani, Paolo and Berti, Emanuele and Gualtieri, Leonardo",
    title = "{Scalar, Electromagnetic and Gravitational Perturbations of Kerr-Newman Black Holes in the Slow-Rotation Limit}",
    eprint = "1307.7315",
    archivePrefix = "arXiv",
    primaryClass = "gr-qc",
    doi = "10.1103/PhysRevD.88.064048",
    journal = "Phys. Rev. D",
    volume = "88",
    pages = "064048",
    year = "2013"
}

@article{Pani:2013pma,
    author = "Pani, Paolo",
    editor = "Cardoso, V. and Gualtieri, L. and Herdeiro, C. and Sperhake, U.",
    title = "{Advanced Methods in Black-Hole Perturbation Theory}",
    eprint = "1305.6759",
    archivePrefix = "arXiv",
    primaryClass = "gr-qc",
    doi = "10.1142/S0217751X13400186",
    journal = "Int. J. Mod. Phys. A",
    volume = "28",
    pages = "1340018",
    year = "2013"
}

@article{SchutzWill1985,
  author = {Schutz, Bernard F. and Will, Clifford M.},
  title = {Black hole normal modes: A semianalytic approach},
  journal = {The Astrophysical Journal},
  volume = {291},
  pages = {L33--L36},
  year = {1985},
  doi = {10.1086/184458},
}

@article{Jansen:2017oag,
    author = "Jansen, Aron",
    title = "{Overdamped modes in Schwarzschild-de Sitter and a Mathematica package for the numerical computation of quasinormal modes}",
    eprint = "1709.09178",
    archivePrefix = "arXiv",
    primaryClass = "gr-qc",
    doi = "10.1140/epjp/i2017-11825-9",
    journal = "Eur. Phys. J. Plus",
    volume = "132",
    number = "12",
    pages = "546",
    year = "2017"
}

@article{IyerWill1987,
  author = {Iyer, Sai and Will, Clifford M.},
  title = {Black-hole normal modes: A {WKB} approach. {I}. Foundations and application of a higher-order {WKB} analysis of potential-barrier scattering},
  journal = {Physical Review D},
  volume = {35},
  number = {12},
  pages = {3621--3631},
  year = {1987},
  doi = {10.1103/PhysRevD.35.3621},
}

@article{Leaver1985,
  author = {Leaver, E. W.},
  title = {An analytic representation for the quasi-normal modes of {Kerr} black holes},
  journal = {Proceedings of the Royal Society of London A},
  volume = {402},
  number = {1823},
  pages = {285--298},
  year = {1985},
  doi = {10.1098/rspa.1985.0119},
}

@article{Cho:2009cj,
    author = "Cho, H. T. and Cornell, A. S. and Doukas, Jason and Naylor, Wade",
    title = "{Black hole quasinormal modes using the asymptotic iteration method}",
    eprint = "0912.2740",
    archivePrefix = "arXiv",
    primaryClass = "gr-qc",
    reportNumber = "YITP-09-110, WITS-CTP-049",
    doi = "10.1088/0264-9381/27/15/155004",
    journal = "Class. Quant. Grav.",
    volume = "27",
    pages = "155004",
    year = "2010"
}

@article{Cho:2011sf,
    author = "Cho, H. T. and Cornell, A. S. and Doukas, Jason and Huang, T. R. and Naylor, Wade",
    title = "{A New Approach to Black Hole Quasinormal Modes: A Review of the Asymptotic Iteration Method}",
    eprint = "1111.5024",
    archivePrefix = "arXiv",
    primaryClass = "gr-qc",
    reportNumber = "YITP-11-97, WITS-CTP-83, OU-HET-735-2011",
    doi = "10.1155/2012/281705",
    journal = "Adv. Math. Phys.",
    volume = "2012",
    pages = "281705",
    year = "2012"
}

@article{ciftci2005construction,
  title={Construction of exact solutions to eigenvalue problems by the asymptotic iteration method},
  author={Ciftci, Hakan and Hall, Richard L and Saad, Nasser},
  journal={Journal of Physics A: Mathematical and General},
  volume={38},
  number={5},
  pages={1147--1155},
  year={2005}
}

@article{barakat2006asymptotic,
  title={The asymptotic iteration method for Dirac and Klein--Gordon equations with a linear scalar potential},
  author={Barakat, T},
  journal={International Journal of Modern Physics A},
  volume={21},
  number={19n20},
  pages={4127--4135},
  year={2006},
  publisher={World Scientific}
}

@article{Hernandez-Velazquez:2021zoh,
    author = "Hernandez-Velazquez, M. I. and Lopez-Ortega, A.",
    title = "{Quasinormal Frequencies of a Two-Dimensional Asymptotically Anti-de Sitter Black Hole of the Dilaton Gravity Theory}",
    eprint = "2108.09559",
    archivePrefix = "arXiv",
    primaryClass = "gr-qc",
    doi = "10.3389/fspas.2021.713422",
    journal = "Front. Astron. Space Sci.",
    volume = "8",
    pages = "713422",
    year = "2021"
}

@article{Blazquez-Salcedo:2016enn,
    author = "Bl{\'a}zquez-Salcedo, Jose Luis and Macedo, Caio F. B. and Cardoso, Vitor and Ferrari, Valeria and Gualtieri, Leonardo and Khoo, Fech Scen and Kunz, Jutta and Pani, Paolo",
    title = "{Perturbed black holes in Einstein-dilaton-Gauss-Bonnet gravity: Stability, ringdown, and gravitational-wave emission}",
    eprint = "1609.01286",
    archivePrefix = "arXiv",
    primaryClass = "gr-qc",
    doi = "10.1103/PhysRevD.94.104024",
    journal = "Phys. Rev. D",
    volume = "94",
    number = "10",
    pages = "104024",
    year = "2016"
}

@article{Blazquez-Salcedo:2017txk,
    author = "Bl{\'a}zquez-Salcedo, Jose Luis and Khoo, Fech Scen and Kunz, Jutta",
    title = "{Quasinormal modes of Einstein-Gauss-Bonnet-dilaton black holes}",
    eprint = "1706.03262",
    archivePrefix = "arXiv",
    primaryClass = "gr-qc",
    doi = "10.1103/PhysRevD.96.064008",
    journal = "Phys. Rev. D",
    volume = "96",
    number = "6",
    pages = "064008",
    year = "2017"
}

@article{Cano:2024ezp,
    author = {Cano, Pablo A. and Capuano, Lodovico and Franchini, Nicola and Maenaut, Simon and V{\"o}lkel, Sebastian H.},
    title = "{Higher-derivative corrections to the Kerr quasinormal mode spectrum}",
    eprint = "2409.04517",
    archivePrefix = "arXiv",
    primaryClass = "gr-qc",
    doi = "10.1103/PhysRevD.110.124057",
    journal = "Phys. Rev. D",
    volume = "110",
    number = "12",
    pages = "124057",
    year = "2024"
}

@article{Brito:2018hjh,
    author = "Brito, Richard and Pacilio, Costantino",
    title = "{Quasinormal modes of weakly charged Einstein-Maxwell-dilaton black holes}",
    eprint = "1807.09081",
    archivePrefix = "arXiv",
    primaryClass = "gr-qc",
    doi = "10.1103/PhysRevD.98.104042",
    journal = "Phys. Rev. D",
    volume = "98",
    number = "10",
    pages = "104042",
    year = "2018"
}

@article{Nomura:2021efi,
    author = "Nomura, Kimihiro and Yoshida, Daisuke",
    title = "{Quasinormal modes of charged black holes with corrections from nonlinear electrodynamics}",
    eprint = "2111.06273",
    archivePrefix = "arXiv",
    primaryClass = "gr-qc",
    reportNumber = "KOBE-COSMO-21-16",
    doi = "10.1103/PhysRevD.105.044006",
    journal = "Phys. Rev. D",
    volume = "105",
    number = "4",
    pages = "044006",
    year = "2022"
}

@article{Cardoso:2018ptl,
    author = "Cardoso, Vitor and Kimura, Masashi and Maselli, Andrea and Senatore, Leonardo",
    title = "{Black Holes in an Effective Field Theory Extension of General Relativity}",
    eprint = "1808.08962",
    archivePrefix = "arXiv",
    primaryClass = "gr-qc",
    doi = "10.1103/PhysRevLett.121.251105",
    journal = "Phys. Rev. Lett.",
    volume = "121",
    number = "25",
    pages = "251105",
    year = "2018",
    note = "[Erratum: Phys.Rev.Lett. 131, 109903 (2023)]"
}

@article{Chung:2024vaf,
    author = "Chung, Adrian Ka-Wai and Yunes, Nicolas",
    title = "{Quasinormal mode frequencies and gravitational perturbations of black holes with any subextremal spin in modified gravity through METRICS: The scalar-Gauss-Bonnet gravity case}",
    eprint = "2406.11986",
    archivePrefix = "arXiv",
    primaryClass = "gr-qc",
    doi = "10.1103/PhysRevD.110.064019",
    journal = "Phys. Rev. D",
    volume = "110",
    number = "6",
    pages = "064019",
    year = "2024"
}

@article{Mamani2022,
  author = {Mamani, Luis A. H. and Masa, Angel D. D. and Sanches, Lucas Timotheo and Zanchin, Vilson T.},
  title = {Revisiting the quasinormal modes of the Schwarzschild black hole: Numerical analysis},
  journal = {Physical Review D},
  volume = {106},
  number = {8},
  pages = {084055},
  year = {2022},
  doi = {10.1103/PhysRevD.106.084055},
  eprint = {2206.03512},
  archivePrefix = {arXiv},
  primaryClass = {gr-qc},
}

@article{Shi2025,
  author = {Shi, Qi-Long and Wang, Rui and Xiong, Wei and Li, Peng-Cheng},
  title = {Quasinormal modes and grey-body factors of axial gravitational perturbations of regular black holes in asymptotically safe gravity},
  journal = {arXiv e-prints},
  pages = {arXiv:2506.16217},
  year = {2025},
  eprint = {2506.16217},
  archivePrefix = {arXiv},
  primaryClass = {gr-qc},
}

@article{Chandrasekhar1975,
  author = {Chandrasekhar, S.},
  title = {On the equations governing the perturbations of the {Schwarzschild} black hole},
  journal = {Proceedings of the Royal Society of London. A. Mathematical and Physical Sciences},
  volume = {343},
  number = {1634},
  pages = {289-298},
  year = {1975},
  publisher = {The Royal Society},
  doi = {10.1098/rspa.1975.0066},
  url = {https://doi.org/10.1098/rspa.1975.0066}
}

@article{ChandrasekharDetweiler1975,
  author = {Chandrasekhar, S. and Detweiler, S.},
  title = {The quasi-normal modes of the {Schwarzschild} black hole},
  journal = {Proceedings of the Royal Society of London. A. Mathematical and Physical Sciences},
  volume = {344},
  number = {1639},
  pages = {441-452},
  year = {1975},
  publisher = {The Royal Society},
  doi = {10.1098/rspa.1975.0034},
  url = {https://doi.org/10.1098/rspa.1975.0034}
}

\end{document}